\newlength{\intwidth}

\documentclass{jfm}

\usepackage{amssymb}

\usepackage{array}
\usepackage{amsmath}
\usepackage{latexsym}
\usepackage{bezier}
\usepackage{graphicx}
\usepackage{epstopdf, epsfig}
\usepackage{psfrag,graphicx}
\usepackage{amsmath,graphicx}
\usepackage{bm}
\usepackage{float}
\usepackage{multirow}
\usepackage{mathrsfs}
\usepackage{color}
\usepackage{enumerate}
\usepackage{natbib}
\usepackage{nicefrac}

\newcommand{\ri}{\mathop{\rm i}\nolimits}
\newcommand{\re}{\mathop{\rm e}\nolimits}

\begin{document}

\title[Instability in centrifugally stable shear flows]{Instability in centrifugally stable shear flows
}

\author{
  Kengo Deguchi\aff{1}\corresp{\email{kengo.deguchi@monash.edu}}
  \and  Ming Dong\aff{2}\corresp{\email{dongming@imech.ac.cn}}
}
 
 \affiliation{
   \aff{1}School of Mathematics, Monash University, VIC 3800, Australia
   \aff{2}State Key Laboratory of Nonlinear Mechanics, Institute of Mechanics, Chinese Academy of Sciences, Beijing 100190, PR China
}

\maketitle

\begin{abstract}
We investigate the linear instability of flows that are stable according to Rayleigh's criterion for rotating fluids. Using Taylor-Couette flow as a primary test case, we develop large Reynolds number matched asymptotic expansion theories. Our theoretical results not only aid in detecting instabilities previously reported by \cite{Deguchi2017} across a wide parameter range 
but also clarify the physical mechanisms behind this counterintuitive phenomenon. Instability arises from the interaction between large-scale inviscid vortices and the viscous flow structure near the wall, which is analogous to Tollmien-Schlichting waves. Furthermore, our asymptotic theories and numerical computations reveal that similar instability mechanisms occur in boundary layer flows over convex walls.



\end{abstract}

\section{Introduction}

Flows deemed unstable by Rayleigh's stability criterion of rotating fluid flows \citep{Rayleigh1917} are often simply referred to as centrifugally unstable flows. Taylor-G\"ortler vortices are textbook examples of flows generated by such instabilities (e.g. \cite{Drazin_Reid1981}). 

The derivation of the Rayleigh criterion relies on two assumptions: axisymmetry and the inviscid nature of the flow. However, the criterion is known to reasonably align with Navier-Stokes based stability analyses for general perturbations. An illustrative example comes from the stability analysis of Taylor-Couette flow, where the Reynolds numbers $R_i$ and $R_o$ pertain to the rotation speeds of the inner and outer cylinders, respectively. In the $R_o$--$R_i$ parameter plane, the region between two lines becomes unstable according to the Rayleigh criterion. One of these lines, known as the Rayleigh line, accurately predicts the onset of Taylor vortices \citep{Taylor1923,Lewis1928}. The other line corresponds to the case where the inner cylinder is fixed and the outer cylinder is rotated. In \cite{Drazin_Reid1981}, this flow is introduced as an example of a flow that is always linearly stable, along with plane Couette flow and Hagen-Poiseuille flow. 

Near the latter $R_i=0$ line, non-axisymmetric disturbances, which the Rayleigh criterion cannot account for, become most unstable. Nevertheless, subsequent physical scaling arguments \citep{EsGr1996}, detailed numerical eigenvalue analysis \citep{Meseguer2002}, and asymptotic theory \citep{Deguchi2016} all agreed on the conclusion that this type of linear instability cannot extend into the Rayleigh stable parameter region, supporting the belief that diffusion and/or non-axisymmetry act as stabilising mechanisms.

It was therefore completely unexpected when \cite{Deguchi2017} numerically found linear instability across the $R_i=0$ line; following \cite{AyDe20}, we refer to this mode as the D17 mode. This mode, also characterised by its non-axisymmetric nature, appears at very high Reynolds numbers, making numerical computations challenging. Although the instability in the centrifugally stable regime has indeed been detected by multiple independent numerical codes, there are still some opinions questioning its existence, as the underlying physical mechanism remains elusive. The primal aim of this paper is to establish beyond doubt the origin of the D17 mode. Our idea is to employ the matched asymptotic expansion analysis, which has a long history of elucidating various shear flow instabilities. In this approach, the largeness of the Reynolds number is rather a favourable effect for constructing mathematically rational approximate solutions and clarifying the physical instability mechanism.

Not much is known about the properties of the D17 mode, but one notable characteristic is its longer wavelength compared to classical instabilities in Taylor-Couette flow. The mode therefore disregards the short-wavelength type stability criteria \citep{BiGa05,KiMu17}, which apply to non-axisymmetric disturbances. \cite{Deguchi2017} speculated on possible similarities between the D17 mode with Tollmien–Schlichting (TS) waves because they both appears in inviscidly stable shear flows and have long-wavelengths. 
\cite{BrRa23} recently analysed spiral Poiseuille flow and reported that the TS wave instability evolves continuously into the D17 modes. However, while TS waves usually exhibit longer wavelengths with increasing Reynolds numbers, the wavelength of the D17 mode in Taylor-Couette flow remains nearly constant. 
Furthermore, even in parallel flows where linear instability does not occur, the D17 mode can still manifest by incorporating Rayleigh-stable effects.
This is evident in the narrow-gap limit of Taylor-Couette flow where the governing equations closely resemble those for plane Couette flow. The above numerical facts indicate that directly applying the well-known TS wave asymptotic theory developed by \cite{Tollmien29}, \cite{Schlichting33}, and \cite{Lin1955} to the D17 mode is not possible.

If the new asymptotic theory for Taylor-Couette flow proves successful, it could also serve as a guide to demonstrate the existence of similar instabilities in other, more practically significant flows. Of particular interest is the 
boundary layer flow over a curved wall. The G\"ortler  vortex, which occurs when the wall surface is concave, has been intensively studied by many researchers \citep{Gortler1941,Hall1982,Hall1983,Hall1988,Saric1994}. While when the wall is convex, \cite{Floryan86} showed that the flow is inviscidly stable applying Raylegh's theorem. Moreover, local stability analysis based on the Orr-Sommerfeld equation has traditionally indicated no instability in this scenario, thus limiting detailed examination. The recent work by \cite{KaHa18} therefore sought a transition route to turbulence that does not rely on linear modal instabilities of the basic flow. Their research is motivated by experiments of boundary layer flows over a wall with convex and concave regions (e.g. \cite{KaMa88}, \cite{BeSa94}). However, to the authors' knowledge, there are no detailed experimental studies focusing exclusively on convex surfaces.

The rest of the paper is structured as follows. In Section 2, we study the narrow-gap limit of Taylor-Couette flow. In this limit, the eigenvalue problem governing stability is greatly simplified, making it ideal for presenting the essence of the asymptotic structure. Our analysis will reveal both similarities and differences between the TS and D17 modes. Section 3 extends the theoretical results to fully wide-gap Taylor-Couette flow. In Section 4, we demonstrate that a similar theory can be applied to boundary layer flow over a convex wall. Finally, in Section 5, we draw some conclusions.



\section{Narrow gap Taylor-Couette flow}

\subsection{Formulation of the problem}
Consider fluid flow between two co-axial cylinders governed by the non-dimensional Navier-Stokes equations 
\begin{eqnarray}\label{NS}
  \partial_{t}\mathbf{v} + (\mathbf{v}\cdot\nabla)\mathbf{v}  =  -\nabla p + \nabla^{2}\mathbf{v}
\end{eqnarray}
and the incompressibility $ \nabla\cdot\mathbf{v}  =  0$. The arrangement of the cylinders is uniquely determined by the radius ratio $\eta=r_i/r_o\leq 1$ when the gap is normalised to be unity. In cylindrical coordinates $(r,\theta,z)$, the no-slip boundary conditions at the cylinder walls are hence prescribed as
\begin{equation}\label{NSBCS}
  \mathbf{v}(r_i,\theta,z)=(0,R_i,0),\quad \mathbf{v}(r_o,\theta,z)=(0,R_o,0), 
\end{equation}
where $r_i=\frac{\eta}{1-\eta}$ and $r_o=\frac{1}{1-\eta}$.
In terms of the dimensional cylinder gap, $d^*$, inner and outer cylinder speeds, $V_i^*,V_o^*$, kinematic viscosity of the fluid, $\nu$, the Reynolds numbers are written as
\begin{equation}
R_i=\frac{V_i^* d^*}{\nu},\qquad R_o=\frac{V_o^* d^*}{\nu}.
\end{equation}
As is well-known, the laminar circular Couette solution can be written as
$(u,v,w)=(0,R_o\,r\Omega,0)$, where the scaled angular velocity has the analytic form
\begin{eqnarray}
\Omega(r)=A+B/r^2 \label{Omega}
\end{eqnarray} 
 with the constants
\begin{eqnarray}
A\equiv \frac{1-\eta^2 a}{1+\eta},\qquad B\equiv \frac{a-1}{1+\eta}r_i^2, \qquad a\equiv \frac{R_i}{\eta R_o}. \label{AB}
\end{eqnarray} 

\subsection{Narrow-gap limit}
\begin{figure}
\begin{center}
\includegraphics[width=0.8\textwidth]{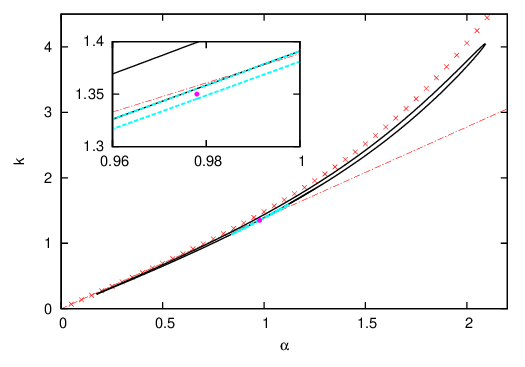}
\end{center}
\caption{
Linear instability found in the narrow-gap Taylor Couette flow problem (\ref{UVeq}) for the stationary inner cylinder case $a=0$. 
The magenta bullet indicate the wavenumber pair $(\alpha,k)=(0.978,1.350)$ at which the instability first emerges at $Re=2.831\times 10^7$. 
The cyan dashed and black solid curves are the neutral curves for $Re=2.9 \times 10^7$ and $10^8$, respectively. The crosses are found by the large $Re$ asymptotic result (\ref{MMM}). 
The dot-dashed line is $k=\frac{5\pi}{8\sqrt{2}}\alpha$ found by the analysis of the Kummer's function. 
}
\label{figak}
\end{figure}

\begin{figure}
\begin{center}
\hspace{-60mm}(a)\\
\includegraphics[width=0.8\textwidth]{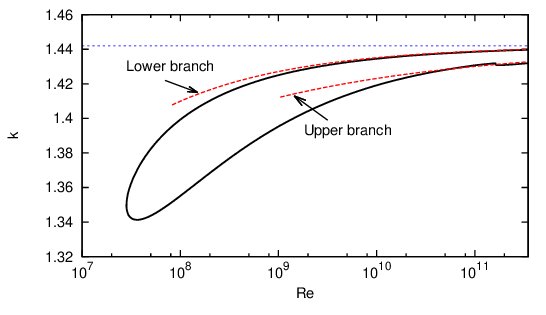}\\
\hspace{-60mm}(b)\\
\includegraphics[width=0.8\textwidth]{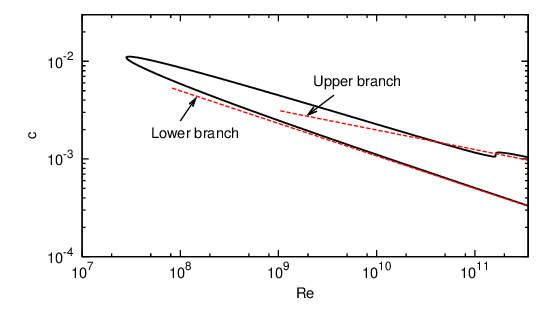}
\end{center}
\caption{
(a) Neutral curve of the narrow-gap Taylor Couette flow problem (\ref{UVeq}) for $\alpha=0.978$, $a=0$ and (b) the corresponding phase speed $c$. 
The full numerical stability results are shown by the black solid curves. The dashed curves are the large $Re$ asymptotic results (\ref{clower}), (\ref{klower}), and (\ref{ckupper}). The blue dotted line in (a) is the leading order approximation $k=1.442$ obtained by (\ref{MMM}). The numerical resolution is verified using up to 3000 Chebyshev modes.
}
\label{figck}
\end{figure}

\begin{figure}
\begin{center}
    \begin{tabular}{cc}                                                         
    \hspace{-40mm}  (a) & \hspace{-40mm} (b) \\
\includegraphics[width=0.47\textwidth]{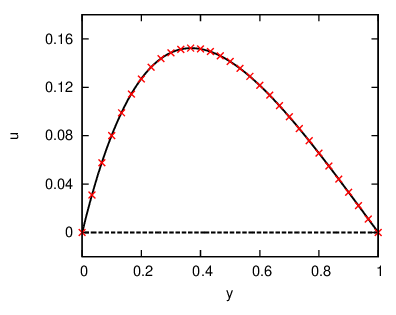}&
\includegraphics[width=0.47\textwidth]{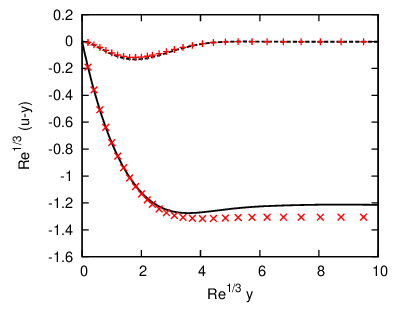}
   \end{tabular} 
\end{center}
\caption{
Comparison of the eigenfunctions between the linearised Navier-Stokes computations (lines) and asymptotic predictions (symbols).
(a) Comparison in terms of the core scaling. The solid and dashed curves are the real and imaginary parts of the lower branch neutral eigenfunction at $(Re,\alpha,k,a)=(10^{10},0.978,1.434,0)$. The points are the leading order core solution $u_0$ with $(\alpha,k_0)=(0.978,1.442)$; see (\ref{u0}). Note that $u_0$ attains its maximum value of $0.1523$ at $y=0.3698$, and $u$ is normalised to have the same property. (b) Comparison in terms of the viscous wall layer scaling. The points are $U_0-Y$, where $U_0$ is the leading order wall layer solution (\ref{U0Airy}) with $(\alpha,k_0,k_1,c_1)=(0.978,1.442,-14.9,2.31)$ and $Y$ is the stretched coordinate $Y=Re^{1/3}y$. Note that the eigenfunction is neutral, and thus the real part of the asymptotic solution approaches $U_d$ as $Y\rightarrow \infty$.
}
\label{figefunc}
\end{figure}

Now we derive the reduced problem valid at the narrow-gap limit $\eta\rightarrow 1$. We introduce a small perturbation parameter $\delta=\sqrt{1-\eta}$ and expand
\begin{eqnarray}
(u,v,w,p)=(u^+, \delta^{-1}v^+,w^+,p^+)+\cdots, ~~ (R_o,R_i)=\delta^{-1}(Re,aRe)+\cdots.
\end{eqnarray}
Changing the coordinates as $y=r-r_i$, $x=\delta^{-1}\theta$ and only retaining the leading-order terms, we get
\begin{subequations}\label{BRE}
\begin{eqnarray}
(\partial_t+v^+\partial_x+u^+\partial_y+w^+\partial_z)u^+-v^{+2}=-p^+_y+(\partial_y^2+\partial_z^2) u^+,\\
(\partial_t+v^+\partial_x+u^+\partial_y+w^+\partial_z)v^+=(\partial_y^2+\partial_z^2) v^+,\\
(\partial_t+v^+\partial_x+u^+\partial_y+w^+\partial_z)w^+=-p^+_z+(\partial_y^2+\partial_z^2) w^+,\\
\partial_x v^+ +\partial_y u^+ +\partial_z  w^+=0,
\end{eqnarray} 
\end{subequations}
and the boundary conditions 
\begin{subequations}
\begin{eqnarray}
(u^+,v^+,w^+)=(0,Re,0)\qquad \text{at}\qquad y=1,\\
(u^+,v^+,w^+)=(0,aRe,0)\qquad \text{at}\qquad y=0.
\end{eqnarray}
\end{subequations}
The limiting process here is essentially that used by \cite{Hall1982, Hall1983, Hall1988} for the G\"ortler vortex problem, and the final set of equations is often referred to as the boundary region equations. The laminar circular Couette flow now becomes a linear profile
\begin{eqnarray}\label{baseNG}
(u,v,w)=(0,Re \, v_b,0),\qquad v_b(y)=(1 - a)y+a.
\end{eqnarray}
Stability of this basic flow to infinitesimal perturbations can be analysed by linearising the equations (\ref{BRE}) and writing $(u^+,v^+,w^+,p^+)=(\hat{u},\hat{v},\hat{w},Re\, \hat{p})\exp(i\alpha (x-Re \, ct)+ikz)$. To ease notation we shall drop the hats after substitution.
Eliminating the pressure and the axial velocity component, we arrive at the coupled equations
\begin{eqnarray}\label{UVeq}
\mathcal{L}v+{{v_b'}} u=0,\qquad
\mathcal{L}( u''-k^2u)+2v_b k^2v=0,
\end{eqnarray}
involving the linear operator $\mathcal{L}=i\alpha (v_b-c)-Re^{-1}(\partial_y^2-k^2)$. The prime denotes ordinary differentiation. 
The above equations and the boundary conditions $u=u'=v=0$ imposed at $y=0$ and 1 constitute an eigenvalue problem for the complex growth rate $c=c_r+ic_i$. 
From the symmetry of the system, it is sufficient to consider positive scaled azimuthal wavenumber $\alpha $ and axial wavenumber $k$. The flow is unstable if $c_i>0$. 

The following mathematical facts are immediately apparent.
\begin{itemize}
\item Squire's theorem does not hold.
\item If $k=0$ the stability problem is equivalent to that for plane Couette flow. Therefore, the well-known proof by \cite{Romanov1973} can be applied, showing that no instability exists.
\item The parameter range $a\in [0,1]$ corresponds to the Rayleigh stable case. In this regime, the flow is stable if $\alpha=0$, i.e. axisymmetric. 
As remarked in section 1, the Rayleigh criterion assumes the inviscid property of the flow. However, even if we include the viscous terms, we can still show the absence of instability, as the proof by \cite{Synge1938} for Taylor-Couette flow is still valid for the narrow-gap limit equations.
\end{itemize}
~\\
It follows that for the Rayleigh stable case $a\in [0,1]$, instability is impossible if either $\alpha$ or $k$ is zero.

In the rest of section 2, we focus on the stationary inner cylinder case $a=0$ for simplicity. The eigenvalue problem (\ref{UVeq}) can be numerically solved by the Chebyshev collocation method. With increasing $Re$ from zero, an instability emerges at $(Re,\alpha,k)=(2.831\times 10^7,0.978,1.350)$, as reported in \cite{Deguchi2017}; see figure \ref{figak}. Like many other shear flows, the size of the instability region in the wavenumber plane expands as the Reynolds number increases further. However, in the current eigenvalue problem, the flow becomes unstable only within a very narrow region, making detection of this instability through numerical computations extremely difficult.

Figure \ref{figck}a shows the neutral curve for $\alpha = 0.978$, with the corresponding phase speed depicted in figure \ref{figck}b. The two branches of the neutral curve emanating from the aforementioned critical point exhibit large Reynolds number asymptotic characteristics somewhat similar to TS waves. Indeed, we shall see in sections 2.3 and 2.4 that the asymptotic structure of the perturbation near the wall bears similarities to that studied in \cite{Tollmien29}, \cite{Schlichting33}, and \cite{Lin45,Lin1955}. For this reason, we adopt the terms upper and lower branches in figure 2 so that they preserve this analogy. 
The red dashed curves in the figures are the asymptotic results to be derived in sections 2.3 and 2.4. For the lower branch, a reasonably accurate asymptotic approximation is obtained for a fairly wide range of $Re$. In contrast, the upper branch exhibits a wiggle, and only beyond this point does the asymptotic result provide a good approximation. These features were also observed in neutral curves of TS waves in channel flows and boundary layer flows (see \cite{Drazin_Reid1981}, for example). 

The magnitude of the phase speed $c$ drops with increasing $Re$ as seen in figure \ref{figck}b, suggesting that if there is a critical layer, it should be sitting around $y=0$, where the base-flow velocity is small. The radial eigenfunction $u$ for the lower branch, shown in figure \ref{figefunc}, indicates that there is indeed some structure near the inner cylinder, while the perturbation does not diminish in the core region away from the wall surface. The nature of the interaction between the core flow and the near-wall layer will naturally be revealed by the method of matched asymptotic expansion. Our theory will also demonstrate that constant wavelengths are necessary to maintain the core structure. Thus, as seen in figure \ref{figck}a, the wavenumber does not decrease with increasing $Re$, unlike TS waves.





\subsection{Lower branch asymptotic analysis}
\label{sec:Lower_narrow_TCF}

\begin{figure}
\begin{center}
\includegraphics[width=1.\textwidth]{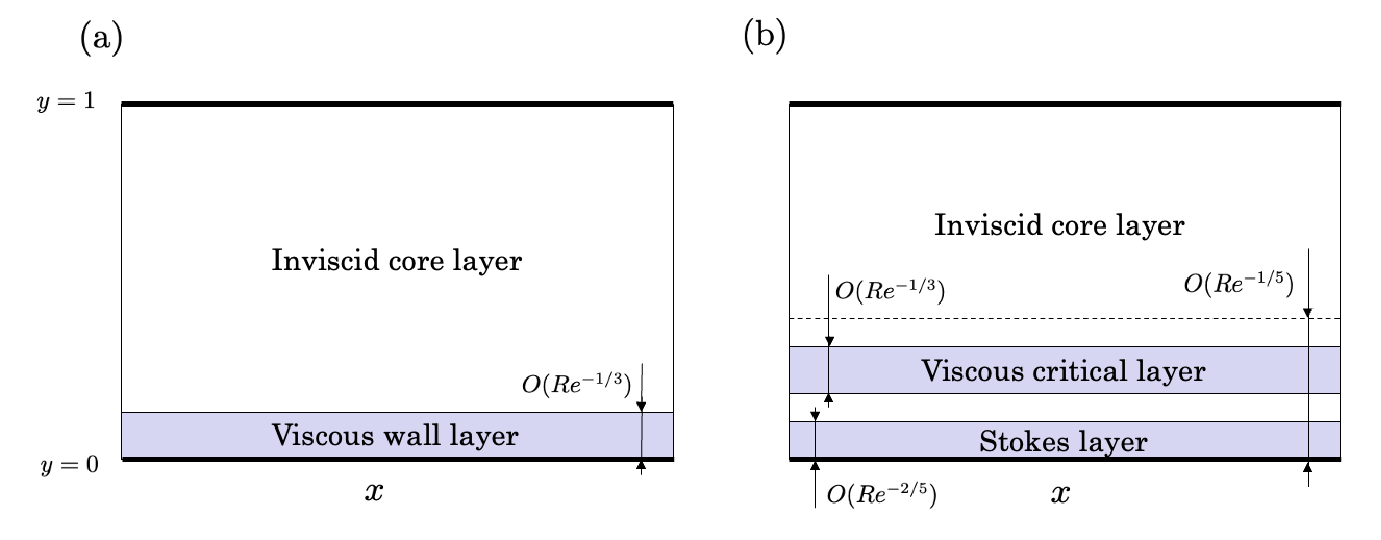}
\end{center}
\caption{
Sketch of large $Re$ asymptotic structure of (a) the lower branch type, and (b) the upper branch type. A Stokes layer of thickness $O(Re^{-1/2})$ exists near $y=1$, but it is omitted.
}
\label{figsketch}
\end{figure}

Let us start by analysing the lower branch, which has a relatively simple asymptotic structure. We fix $\alpha$ at a positive constant throughout the analysis. When $Re$ is asymptotically large, the majority of the fluid layer is occupied by an inviscid core region, while near the inner cylinder ($y=0$), a thin viscous wall layer of thickness $O(\epsilon)$ emerges (see figure \ref{figsketch}a). Since we have set $a=0$, within the boundary layer, the size of $v_b$ is $O(\epsilon)$. Therefore, the advective effect of $O(\epsilon)$ and the viscous effect of $O(Re^{-1}\epsilon^{-2})$ in the operator $\mathcal{L}$ in (\ref{UVeq}) balance when $\epsilon=Re^{-1/3}$. This wall layer thickness is typical of critical layers, where the phase speed of the wave matches the base flow speed. 

Using the small parameter $\epsilon$, we expand $k=k_0+\epsilon k_1+\cdots$ and $c=\epsilon c_1 +\cdots$. The overall structure of the asymptotic problem is as follows. First, the leading order inviscid core problem determines $k_0$. Then, at the next order, the core flow is coupled with the wall layer problem to yield the dispersion relation linking $c_1$ and $k_1$. 
In the neutral case, $c_1$ and $k_1$ can be solved explicitly.

In the inviscid core region, we expand the flow field as
\begin{eqnarray}\label{expcore1}
u=u_0(y)+\epsilon u_1(y)+\cdots,\qquad v=v_0(y)+\epsilon v_1(y)+\cdots.
\end{eqnarray}
Substituting (\ref{expcore1}) into (\ref{UVeq}), retaining only the leading order terms, and eliminating $v_0$, we have the single second-order ordinary differential equation
\begin{eqnarray}\label{core0th}
\mathscr{L} u_0 =0, \qquad \mathscr{L}\equiv \partial_y^2+k_0^2 (\frac{2}{\alpha^2  y}-1). \label{v0eq}
\end{eqnarray}
A similar manipulation at the next order yields the inhomogeneous version of the equation
\begin{eqnarray}\label{core1st}
\mathscr{L} u_1 = - \left \{\frac{4 k_0^2c_1}{\alpha^2 y^2}
+2k_1 k_0 (\frac{2}{\alpha^2y}-1)\right \} u_0.\label{v1eq}
\end{eqnarray}
From the no-penetration condition at the wall, we impose $u_0(1)=u_1(1)=u_0(0)=0$. 
However, it will turn out from the boundary layer analysis that $u_1(0)$ cannot be zero, so we denote this displacement velocity as $U_d=u_1(0)$.

The no slip conditions at the outer cylinder ($y=1$) are satisfied through a Stokes layer of thickness $O(Re^{-1/2})$. However, this layer is passive 
and does not need to be analysed.

The leading order core system is an eigenvalue problem for the eigenvalue $k_0^2$.
The solution can be written down using the Kummer's function of the first kind $M$: 
\begin{eqnarray}\label{u0}
u_0(y)=ye^{-k_0 y}\frac{M(\gamma, 2, 2k_0 y)}{M(\gamma, 2, 0)},\qquad \gamma\equiv 1-\frac{k_0}{\alpha^2}.
\end{eqnarray}
The boundary condition at $y=1$ is fulfilled only when
\begin{eqnarray}\label{MMM}
M(\gamma, 2, 2k_0)=0.
\end{eqnarray}
The crosses in figure \ref{figak} are obtained by looking for the smallest positive $k_0$ that satisfies (\ref{MMM}) for each $\gamma$.
For a small $\alpha$, the approximation
\begin{eqnarray}
k_0\approx \frac{5\pi}{8\sqrt{2}}\alpha
\end{eqnarray}
is available, thanks to the property of the Kummer's function \citep{AbSt1965}).
Both results are useful for estimating the wavenumber region where instability might occur.
For $\alpha=0.978$ used in figure \ref{figck}a the prediction by (\ref{MMM}) gives
$k_0\approx 1.442,$
which is shown by the blue dotted line in the figure. 
Also, the leading order core solution (\ref{u0}) agrees very well with the numerically obtained neutral eigenfunction, as depicted in figure \ref{figefunc}a. Note that to simplify the subsequent analysis we have normalised the eigenfunction imposing  $u_0'(0)=1$.

Within the viscous wall layer, we use the stretched variable $Y=y/\epsilon$ and expand
\begin{eqnarray}\label{expwall1}
u=\epsilon U_0(Y)+\cdots, \qquad v=V_0(Y)+\cdots.
\end{eqnarray}
Here, the normalisation $u_0'(0)=1$ and the existence of the displacement velocity $u_1(0)=U_d$ imply that $U_0 \sim Y+U_d$ for large $Y$ to match the boundary layer solution to the core solutions. While at $Y=0$, of course $V_0,U_0,$ and $U_0'$ must vanish to satisfy the no-slip conditions. Substituting (\ref{expwall1}) and the Taylor expansion $v_b=\lambda y+\cdots$ into (\ref{UVeq}), the leading order equations are obtained as
\begin{eqnarray}
(i\alpha  ({\lambda}Y-c_1)-\partial_Y^2)V_0+{\lambda}U_0=0,\label{wallV}\\
(i\alpha  ({\lambda}Y-c_1)-\partial_Y^2)U_0''=0.\label{wallU}
\end{eqnarray}
From (\ref{baseNG}), $\lambda$ is simply unity when $a=0$; however, for our later purposes, we will carry out the calculations while keeping the general $\lambda$. 
The change of the variable $\xi=\sigma Y+\xi_0$ with the constants $\xi_0=-\sigma c_1/{\lambda}$ and $\sigma =(i\lambda \alpha)^{1/3}$ transforms (\ref{wallU}) to Airy's equation for $\partial_{\xi}^2U_0$:
\begin{eqnarray}
(\xi-\partial_{\xi}^2)\partial_{\xi}^2U_0=0.
\end{eqnarray}
After some algebra, the solution satisfying the boundary conditions can be found as
\begin{eqnarray}\label{U0Airy}
\hspace{-3mm}
U_0=\sigma^{-1}\left [\xi+\frac{1}{\kappa(\xi_0)}\left (-\xi \kappa(\xi)-\text{Ai}'(\xi)+\text{Ai}'(\xi_0) \right) \right ].
\end{eqnarray}
For large $Y$, the magnitude of the functions $\kappa(\xi)=\int^{\infty}_{\xi}\text{Ai}(\check{\xi})d\check{\xi}$ and $\text{Ai}'(\xi)$ rapidly diminish. Therefore, referring to the aforementioned matching condition, the displacement velocity is found as
\begin{eqnarray}
U_{d}= \frac{\text{Ai}'(\xi_0)}{\sigma \kappa(\xi_0)}-\frac{c_1}{{\lambda}}.
\label{eq:U_infty}
\end{eqnarray}
This is the only information we need for the dispersion relation in the wall layer analysis. Hence, we do not solve (\ref{wallV}) to find $V_0$.

The dispersion relation readily follows from the solvability condition for the inhomogeneous equation (\ref{v1eq}). The adjoint solution is merely $u_0$ because the homogeneous equation (\ref{v0eq}) is self-adjoint. Multiplying (\ref{v1eq}) by $u_0$ and integrating from 0 to 1, 
\begin{eqnarray}\label{core1st}
\hspace{-3mm}
\int^1_0 u_0 \mathscr{L} u_1 dy= - \int^1_0 \left \{\frac{4 k_0^2c_1}{\alpha^2 y^2}
+2k_1 k_0 (\frac{2}{\alpha^2y}-1)\right \} u_0^2 dy.
\end{eqnarray}
The left-hand side becomes $U_d$ after integration by parts and using (\ref{v0eq}). Therefore, $k_1$ and $c_1$ are now linked by
\begin{eqnarray}
-\frac{1}{(\lambda \alpha)^{1/3}} \frac{i^{5/3}\text{Ai}'(\xi_0)}{ \kappa(\xi_0)}+{c_1} (I_1-\frac1{{\lambda}})+k_1I_2 =0,\label{I1I2}
\end{eqnarray}
where 
\begin{eqnarray}\label{intI1I2}
I_1(\alpha)=\frac{4 k_0^2}{\alpha^2}
\int^1_0 
\frac{u_0^2}{y^2} dy,\qquad
I_2(\alpha)=2k_0 \int^1_0 (\frac{2}{\alpha^2y}-1)u_0^2 dy,
\end{eqnarray}
and $\xi_0=i^{7/3}(\lambda \alpha)^{1/3} c_1/{{\lambda}}$.

If the mode is neutral (i.e. $c_1$ is purely real), the imaginary part of (\ref{I1I2})  yields
\begin{eqnarray}
\Im \left \{ \frac{i^{5/3}\text{Ai}'(i^{7/3}s)}{ \kappa(i^{7/3}s)} \right \}=0,\qquad s=(\lambda \alpha)^{1/3} c_1/{{\lambda}}.
\end{eqnarray}
A real root of this function can be found at $s=s_0\approx 2.2972$ \citep{Miles1960}. 
Setting $\lambda=1$, the following approximation is obtained for the lower branch shown in figure \ref{figck}b:
\begin{eqnarray}\label{clower}
c=2.2972(0.978Re)^{-1/3}.
\end{eqnarray}
The real part of (\ref{I1I2}) can now be solved for $k_1$. 
\begin{eqnarray}\label{k1int}
k_1 =\frac{q_0
+s_0 (1-I_1)}{\alpha^{1/3}I_2 },\qquad q_0=\frac{i^{5/3}\text{Ai}'(i^{7/3}s_0)}{ \kappa(i^{7/3}s_0)}\approx 
{-1.00.}
\end{eqnarray}
The integrals (\ref{intI1I2}) can be worked out numerically after substituting $u_0$ found in (\ref{u0}). For $\alpha=0.978$, using $k_0=1.442$ estimated before, the right-hand side of (\ref{k1int}) can be computed as approximately 14.9. Therefore, we have the asymptotic result
\begin{eqnarray}\label{klower}
k=1.442-14.9Re^{-1/3}
\end{eqnarray}
used in figure \ref{figck}a. The wall layer solution (\ref{U0Airy}) is compared with the numerical neutral solution in figure \ref{figefunc}b.

\subsection{Upper branch asymptotic analysis}
The analysis for the upper branch is more complicated, and thus the neutrality of the perturbation is assumed from the beginning to facilitate discussion. 
The critical difference to the lower branch is that the majority of the wall layer becomes inviscid. Viscosity acts within two thinner sublayers: the critical layer sitting in the middle of the inviscid zone and the Stokes layer adjacent to the wall (see figure \ref{figsketch}b). 

The core structure is identical to the lower branch case, but the wall layer thickness, again denoted as $\epsilon$, is different. Within the wall layer, the radial velocity is $O(\epsilon)$ to leading order, and it is merely a linear profile. Since the critical layer appears within the wall layer, the phase speed $c$ must be $O(\epsilon)$. A displacement velocity induced across the critical layer affects the next-order component of the wall layer radial velocity, which is $O(\epsilon^2)$. A displacement velocity is also induced by the Stokes layer, and it must match to the contribution from the critical layer.
The former velocity is proportional to the Stokes layer thickness $\Delta$, so we require $\Delta=\epsilon^2$. Finally, balancing the $O(c)$ advection term and the $O(Re^{-1}\Delta^{-2})$ diffusion term in the Stokes layer, we arrive at the scaling $\epsilon=Re^{-1/5}$ and $\Delta=Re^{-2/5}$. The critical layer thickness $\tilde{\epsilon}=Re^{-1/3}$ is indeed thinner than the wall layer thickness $\epsilon$. 

We begin the formal asymptotic analysis by writing $k=k_0+\epsilon k_1+\cdots$ and $c=\epsilon c_1 +\cdots$. Within the wall layer, the velocity components are expanded as
\begin{eqnarray}
u=\epsilon U_0(Y)+\epsilon^2 U_1(Y)+\cdots,\qquad v=V_0(Y)+\epsilon V_1(Y)+\cdots,
\end{eqnarray}
using the functions of $Y=y/\epsilon$.
Substituting them into (\ref{UVeq}), the leading order equations are obtained as
\begin{subequations}
\begin{eqnarray}
i\alpha ( Y- c_1)V_0+U_0=0,\\
i\alpha  ( Y- c_1) U''_0=0,
\end{eqnarray}
\end{subequations}
and the next order equations are
\begin{subequations}
\begin{eqnarray}
i\alpha ( Y- c_1)V_1+U_1=0,\\
i\alpha  ( Y- c_1) U_1''+2Yk_0^2V_0=0.\label{U1equpper}
\end{eqnarray}
\end{subequations}
For $U_0$, we impose $U_0(0)=0$ to satisfy the non-penetrating condition on the wall. 
Also, $U_0\sim Y+U_{d}$ as $Y\rightarrow \infty$ to match the normalised core solution. 
Then the solution to the leading order problem can be found as
\begin{eqnarray}\label{U0upper}
U_0=Y, \qquad V_0=-\frac{Y}{i\alpha( Y- c_1)},
\end{eqnarray}
which yields $U_{d}=0$.
The singularity that appears in $V_0$ should be resolved within the critical layer around $Y=c_1$.
Given $V_0$, we can integrate (\ref{U1equpper}) twice to find
\begin{eqnarray}\label{UU1upper}
 U_1=-\frac{k_0^2}{\alpha^2}\left \{(Y-c_1)^2
 + 4c_1( Y- c_1)[\ln |Y- c_1|+\phi] \right .\nonumber \\
\left .  -2c_1^2[\ln |Y- c_1|+\psi]
   \right \}.
\end{eqnarray}
The constants $\phi$ and $\psi$ are unknown and may differ for $Y>c_1$ and $Y<c_1$.
To distinguish the constants in the former region from those in the latter region, we attach a subscript $+$ to the former and $-$ to the latter. The constants $\phi_+$ and $\psi_+$ must be purely real to match the core solution which is still inviscid at the corresponding order. The critical layer analysis reveals that 
\begin{eqnarray}\label{phipsi}
\phi_- =\phi_+ -i\pi,\qquad \psi_-=\psi_+ -i\pi.
\end{eqnarray}
This `logarithmic phase shift' may not be surprising to readers familiar with the asymptotic analysis of shear flows. 
To find this shift, one needs to use the expansions 
\begin{subequations}
\begin{eqnarray}
u&=&\epsilon c_1+\tilde{\epsilon} \zeta+O(\epsilon^2\ln \epsilon)+\epsilon^2 \tilde{U}_1(\zeta)+O(\epsilon \tilde{\epsilon} \ln \epsilon)+\epsilon \tilde{\epsilon} \tilde{U}_2(\zeta)+\cdots,\\
v&=&\epsilon \tilde{\epsilon}^{-1} \tilde{V}_0(\zeta)+\tilde{V}_1(\zeta)+\cdots.
\end{eqnarray}
\end{subequations}
valid when $\zeta=\tilde{\epsilon}^{-1}(y- \epsilon c_1)$ is $O(1)$.
Equations (\ref{UU1upper}) and (\ref{phipsi}) imply that 
\begin{eqnarray}\label{shearImu}
\Im(u)\rightarrow -\epsilon^2 \frac{6\pi c_1^2 k_0^2}{\alpha^2} \qquad \text{as}\qquad Y\rightarrow 0.
\end{eqnarray}

Within the Stokes layer, we write $Z=y/\epsilon^2$, expand
\begin{eqnarray}\label{Stexp}
u=\epsilon^2\check{U}_0(Z)+\cdots,\qquad v=\check{V}_0(Z)+\cdots,
\end{eqnarray}
and impose the boundary conditions $\check{V}_0(0)=\check{U}_0(0)=\check{U}'_0(0)=0$.
Substitution of (\ref{Stexp}) to (\ref{UVeq}) yields the leading order equations
\begin{eqnarray}
(i\alpha c_1+\partial_{Z}^2)\check{V}_0=0,\qquad (i\alpha c_1+\partial_{Z}^2)\check{U}_0''=0.
\end{eqnarray}
The solution of the second equation, which does not grow exponentially for large $Z$, can be found as follows:
\begin{eqnarray}
\check{U}_0=Z+\frac{e^{-qZ}-1}{q}, \qquad q=(1-i)\sqrt{\frac{\alpha c_1}{2}}.
\end{eqnarray}
The exponential part is negligible when $Z$ is large.
Thus on exiting the Stokes layer
\begin{eqnarray}\label{StokesImu}
\Im(u)\rightarrow -\epsilon^2 \frac{1}{(2\alpha c_1)^{1/2}} \qquad \text{as}\qquad Z\rightarrow \infty.
\end{eqnarray}
Matching (\ref{shearImu}) and (\ref{StokesImu}), the scaled phase speed is found as
\begin{eqnarray}\label{c1up}
 c_1= \frac{\alpha^{3/5}}{2^{1/5} (6\pi k_0^2)^{2/5}}.
\end{eqnarray}
Here, we can use (\ref{MMM}) to find $k_0$, as the core analysis is identical to that for the lower branch case.
On the other hand, the dispersion relation (\ref{k1int}) simplifies to
\begin{eqnarray}\label{k1intup}
k_1 =-\frac{c_1I_1}{I_2}
\end{eqnarray}
because $u_1(0)=U_{d}=0$ as remarked just below (\ref{U0upper}). 

Substitution of $\alpha=0.978$ and $k_0=1.442$ into (\ref{c1up}) and (\ref{k1intup}) gives $c_1\approx 0.198$ and $k_1\approx 1.89$. 
The asymptotic approximations shown in figure \ref{figck} are thus obtained as
\begin{eqnarray}\label{ckupper}
c=0.198Re^{-1/5},\qquad k=1.442-1.89Re^{-1/5}.
\end{eqnarray}

\section{Wide gap Taylor-Couette flow}

The analysis presented so far can be extended to the full Taylor-Couette flow problem introduced in Section 2.1. 
The radius ratio $\eta$ now takes the value between 0 and 1, and the rotation ratio $a$ is not necessarily zero. 

Linearising the Navier-Stokes equations (\ref{NS}) around the circular Couette flow, substituting $(u,v,w,p)=(\hat{u},\hat{v},\hat{w},R_o\, \hat{p})\exp(im (\theta-R_o \, ct)+ikz)$, and removing the hats, we have the well-known eigenvalue problem:
\refstepcounter{equation}\label{TCFev}
$$
\mathcal{L}u-2\Omega v+\frac{1}{R_o}\Big(\frac{u}{r^2} +\frac{2 i m v}{r^2}\Big)=-p',\eqno{(\theequation a)}
$$
$$
\mathcal{L}v+(2\Omega +r\Omega')u+\frac{1}{R_o}\Big(\frac{v}{r^2}-\frac{2i mu}{r^2}\Big)=-\frac{im}{r}p,\eqno{(\theequation b)}
$$
$$
\mathcal{L}w=-ik{p},\eqno{(\theequation c)}$$
$$
u'+\frac{u}{r}+\frac{im}{r}v+ikw=0.\eqno{(\theequation d)}\label{eq:OS}
$$
The no-slip conditions $u=v=w=0$ are imposed at $r=r_i$ and $r_o$.
To compactly write the equations, we introduced the linear operator $\mathcal{L}=im (\Omega-c)-R_o^{-1}(\partial_r^2+r^{-1}\partial_r-r^{-2}m^2-k^2)$. 

Section 3.1 extends the lower branch type asymptotic analysis deduced in section 2.3 for the narrow-gap limit case. In view of (\ref{eq:OS}), it is clear that the suitable large asymptotic parameter is $R_o$. We only study small enough
$|a|$, so that the non-zero rotation ratio effect preserves the asymptotic structure studied in section 2.3. The angular velocity (\ref{Omega}) of the circular Couette flow hence admits the Taylor expansion $\Omega=\Omega_0+a \Omega_1+\cdots$, where
\begin{equation}
\Omega_0=\frac1{1+\eta}-\frac{r_i^2}{r^2(1+\eta)},\quad \Omega_1=-\frac{\eta^2}{1+\eta}+\frac{r_i^2}{r^2(1+\eta)}.
\end{equation}
In section 3.2, the asymptotic results are compared with the full numerical solution of (\ref{eq:OS}).

The extension of the upper branch theory in section 2.4 to the wide-gap case is straightforward, but given its limited practical importance, we omit this discussion.

\subsection{Asymptotic analysis}

We introduce a small parameter $\epsilon=R_o^{-1/3}$ that describes the thickness of the wall layer near the inner cylinder. Note that this $\epsilon$ is not the same as the one used in section 2.3; we will clarify their relationship later. The rotation rate $a$ is taken to be $O(\epsilon)$ so that $a_1=a/\epsilon$ is of order unity. The azimuthal wavenumber $m$ is held fixed, while the axial wavenumber and the complex phase speed are written in the forms $k=k_0+\epsilon k_1+\cdots$ and $c=\epsilon c_1+\cdots$, respectively.

In the inviscid core region, we expand
\begin{eqnarray}
(u,v,w,p)=(u_0,v_0,w_0,p_0)+\epsilon (u_1,v_1,w_1,p_1)+\cdots.
\end{eqnarray}
The terms appearing in the right hand side are all functions of $r$.
Substitution of those expansions to (\ref{eq:OS}) yields the leading order equation for $u_0(r)$,
\begin{equation}
\mathscr{L}u_0=0,
\label{eq:normal_main}
\end{equation}
and the next order equation for $u_1(r)$,
\begin{equation}
\mathscr{L}u_1=k_1F_2 u_0'-\Big(c_1F_0+
k_1F_1+a_1F_3 \Big)u_0.\label{eq:u_1_wide}
\end{equation}
The differential operator $\mathscr{L}$ and the functions $F_0, F_1, F_2$, and $F_3$ depend on $m,k_0,\eta$ as can be found in Appendix A. The boundary conditions are $u_0(r_i)=u_0(r_o)=u_1(r_o)=0$ and $u_1(r_i)=U_d$, where $U_d$ is the displacement velocity to be found by the wall layer analysis.

Within the wall layer near the inner cylinder, the suitable expansions are
\begin{eqnarray}
(u,v,w,p)=(\epsilon U_0,V_0,W_0,\epsilon P_0)+\cdots,
\end{eqnarray}
where the terms in the right-hand side are functions of $Y=\epsilon^{-1}(r-r_i)$. Within the layer where $Y$ is of order unity, the base-flow angular velocity (\ref{Omega}) behaves like $\Omega=\epsilon (\lambda Y+b a_1 )+\cdots$ with the constants 
\begin{eqnarray}
\lambda\equiv \Omega_0'(0)=\frac{2}{r_i(1+\eta)},\qquad b\equiv  \Omega_1(0)={1-\eta}.
\end{eqnarray}
After some manipulations, from the leading order problem we can find that $U_0(Y)$ satisfies
\begin{eqnarray}
(im  (\lambda Y+ba_1-c_1)-\partial_Y^2)U_0''=0
\end{eqnarray}
which, after re-reading $m, (c_1-ba_1)$ as $\alpha$, $c_1$, coincides with (\ref{wallU}).
Therefore we can recycle the narrow-gap result to obtain the displacement velocity
\begin{eqnarray}\label{UdTCF}
U_d= -\frac{i^{5/3}\text{Ai}'(\xi_0)}{(\lambda m)^{1/3} \kappa(\xi_0)}-\frac{c_1-b a_1}{\lambda},\qquad \xi_0=\frac{i^{7/3}(\lambda m)^{1/3}(c_1-b a_1)}{\lambda}.
\end{eqnarray}

To solve the inhomogeneous problem (\ref{eq:u_1_wide}), we use the inner product $\langle f,g \rangle\equiv \int_{r_i}^{r_o} f g dr$. The homogeneous problem (\ref{eq:normal_main}) is not self-adjoint, so we introduce the adjoint problem
\begin{equation}\label{wideadj}
\mathscr{L}^\dag u^\dag=0, \qquad u^\dag(r_i)=u^\dag(r_o)=0,
\end{equation}
where the operator $\mathscr{L}^\dag$ is given in Appendix A.
The solution $u^\dag(r)$ of this problem with 
the normalisation $(\hat u^\dag)'(r_i)=1$ can be uniquely determined. 
Taking the inner product of $u^\dag$ and (\ref{eq:u_1_wide}), we obtain the dispersion relation
\begin{equation}\label{diswide}
U_d+{\langle u^\dag,F_0 u_0 \rangle}c_1-k_1\langle u^\dag,F_2 u_0'-F_1 u_0\rangle+a_1\langle  u^\dag,F_3 u_0\rangle=0,
\end{equation}
where $U_d(m,\eta,c_1,a_1)$ is given in (\ref{UdTCF}).





\subsection{Numerical analysis}

\begin{figure}
 \begin{center}
\includegraphics[width = 0.48\textwidth] {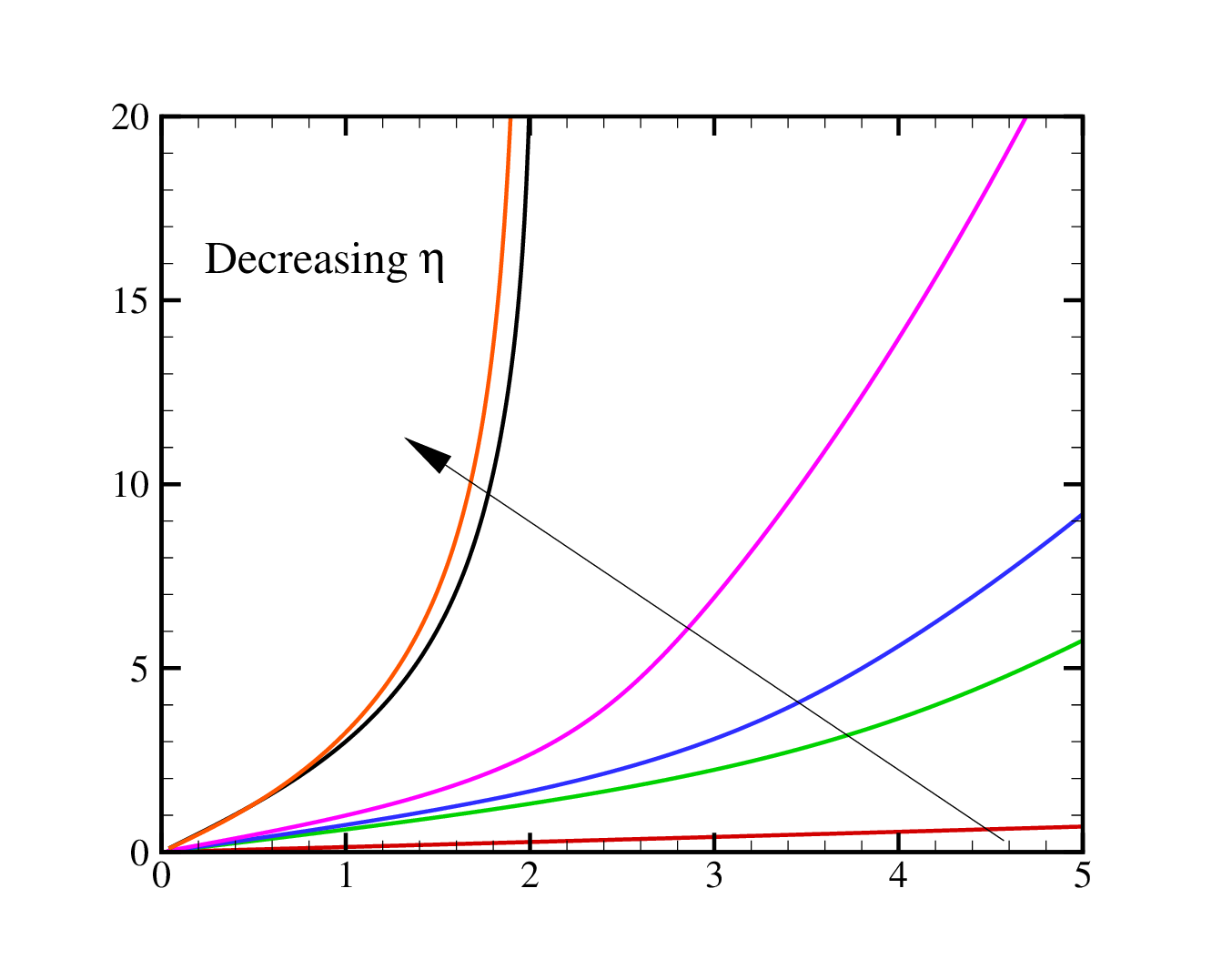}
     \put(-95,1){$m$}\put(-180,70){{$k_0$}} \put(-180,115){(a)}
\includegraphics[width = 0.48\textwidth] {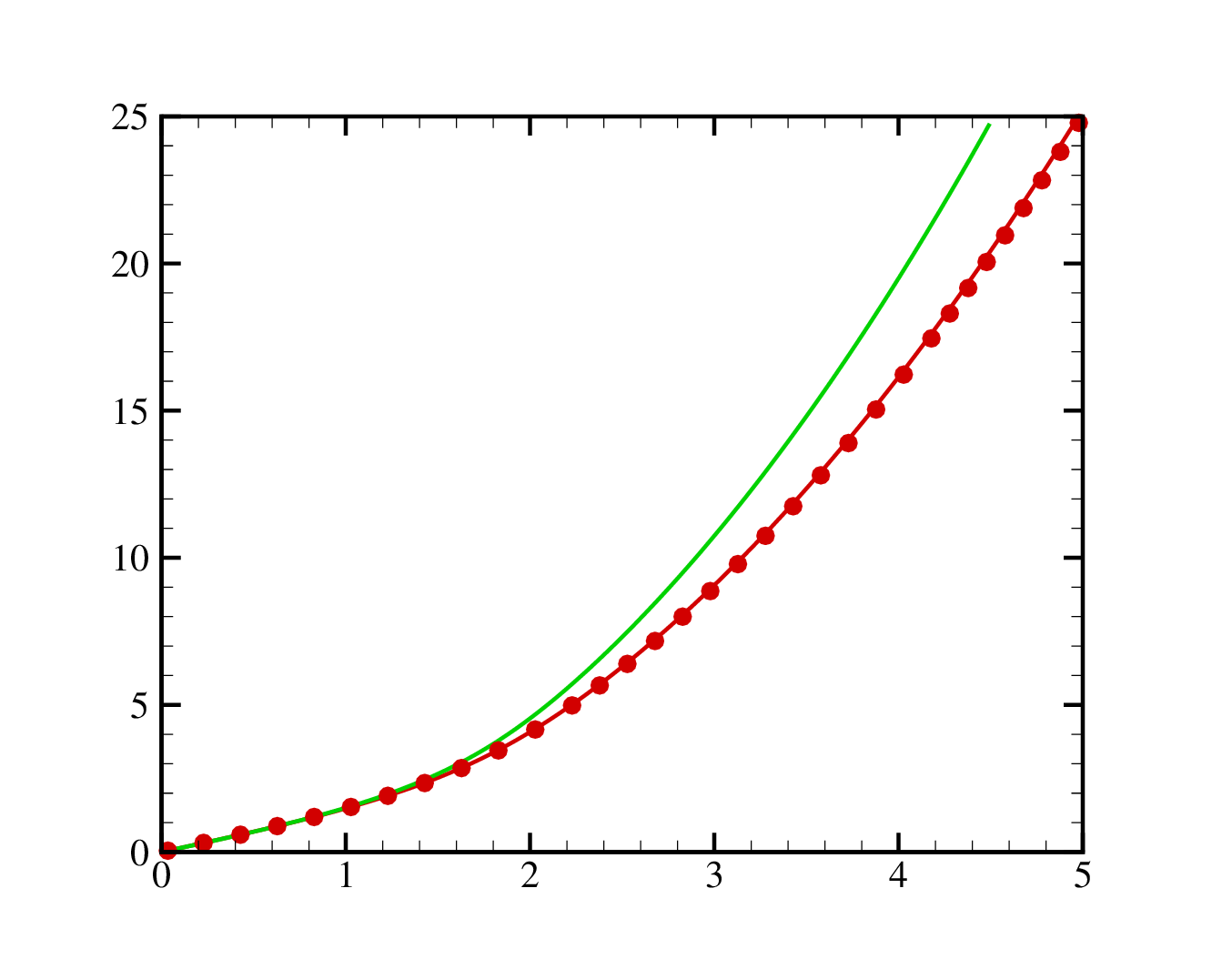}
     \put(-95,1){$\delta m$} \put(-180,70){{$k_0$}}\put(-180,115){(b)}
   \caption{
  Asymptotic results obtained by numerically solving (\ref{eq:normal_main}).
 (a): Dependence of the leading-order axial wavenumber $k_0$ on the azimuthal wavenumber $m$ for $\eta=0.99$ (red), 0.8 (green), 5/7 (blue), 0.5 (pink), 0.1 (black), 0.01 (orange). 
 (b): The red and green curves are the results for $\eta=0.99$ and 0.8, respectively, shown in the $\delta m$--$k_0$ plane. The circles are the narrow-gap limit results assuming $(k_0,\delta m)=(\overline{k}_0,\alpha)$.
    }
\label{fig:comparison_leading}
\end{center}
\end{figure}

\begin{figure}
 \begin{center}
 \includegraphics[width = 0.48\textwidth] {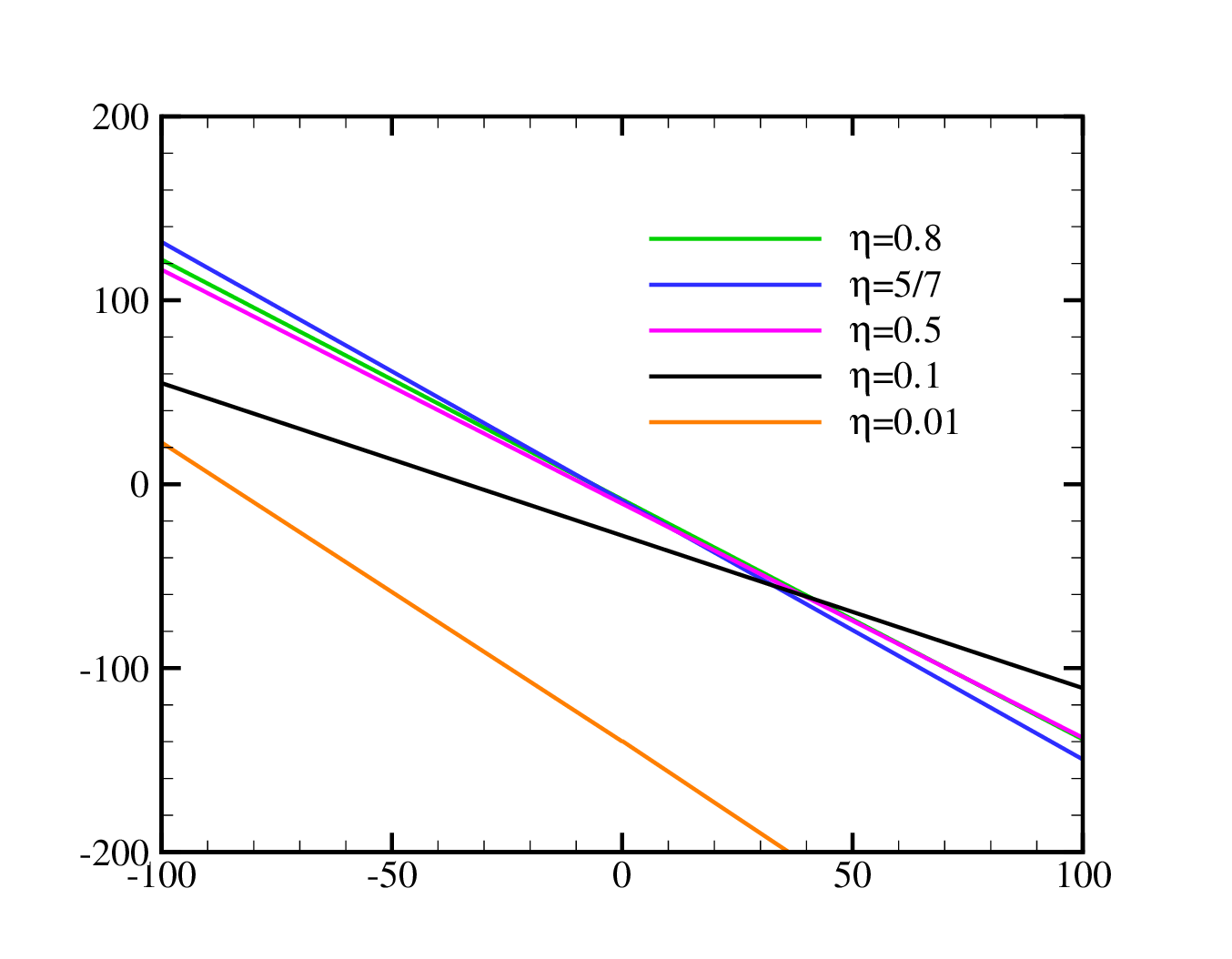}
     \put(-96,1){$a_1$}\put(-180,70){$k_1$}\put(-180,115){(a)}
\includegraphics[width = 0.48\textwidth] {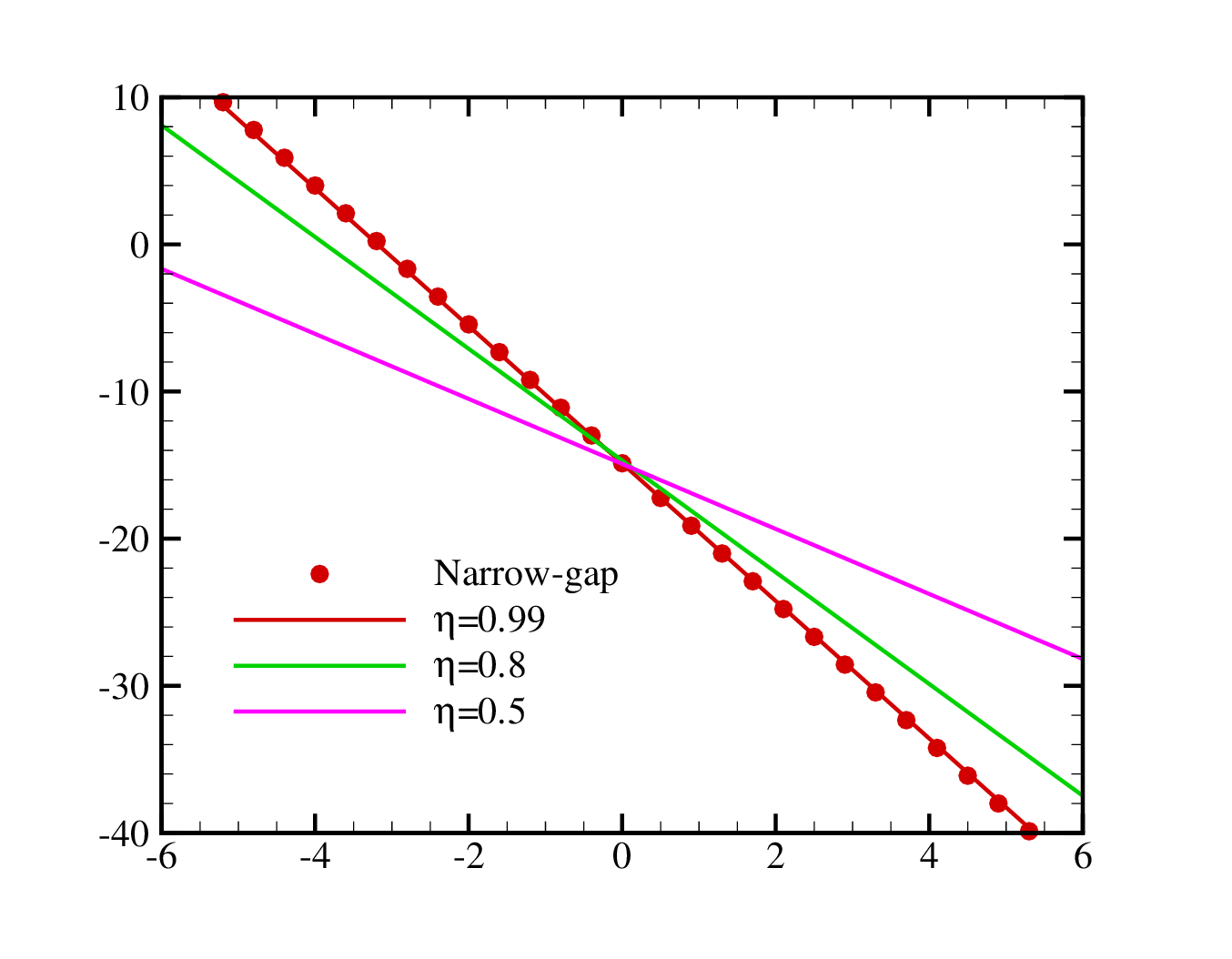}
     \put(-96,1){$\delta^{1/3}a_1$}\put(-190,70){$\delta^{1/3} k_1$}\put(-180,115){(b)}
   \caption{Dependence of $k_1$ on $a_1$ for the neutral modes computed by (\ref{diswide}). 
   (a) The results for wide-gap configurations. The azimuthal wavenumber is fixed at $m=1$. (b) 
   The results for narrow-gap configurations. The azimuthal wavenumber is varied as $m=0.978/\delta$ to observe the convergence to the narrow-gap results with $\alpha=0.978$ (circles).
In the narrow-gap results, we set $(\delta^{1/3}k_1,\delta^{1/3}a_1)=(\overline{k}_1,\overline{a}_1)$. 
 }
\label{fig:comparison2_narrow}
\end{center}
\end{figure}
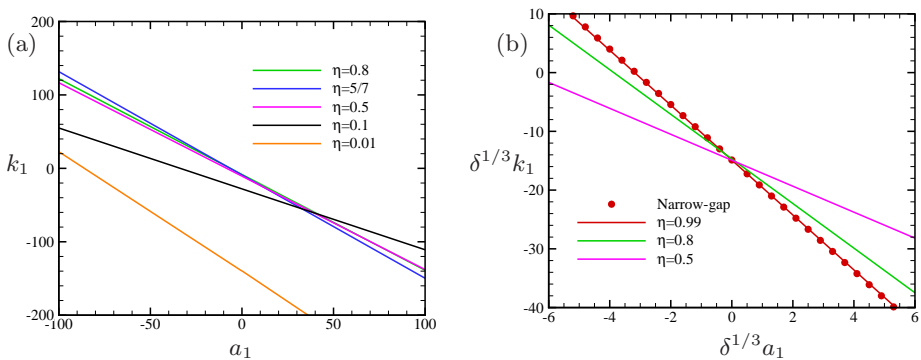
The wide-gap version of the asymptotic results requires numerical work to extract useful information. First, the leading order problem (\ref{eq:normal_main}) needs to be solved. The two boundary conditions can be satisfied only when specific values of $k_0(m,\eta)$ are taken. The admissible values of $k_0$ for various $\eta$ are summarised in figure \ref{fig:comparison_leading}a. Only integer values of $m$ are physically relevant, though real values of $m$ can be used in numerical computation without issue.
If the perturbations are assumed to be neutral, the next step is to find $c_1(m,\eta,a_1)$ from the imaginary part of (\ref{diswide}). Then the real part of (\ref{diswide}) can be used to compute $k_1(m,\eta,a_1)$ after some numerical integrations. The dependence of $k_1$ on $a_1$ is perfectly linear, as shown in figure \ref{fig:comparison2_narrow}a for the case $m=1$.


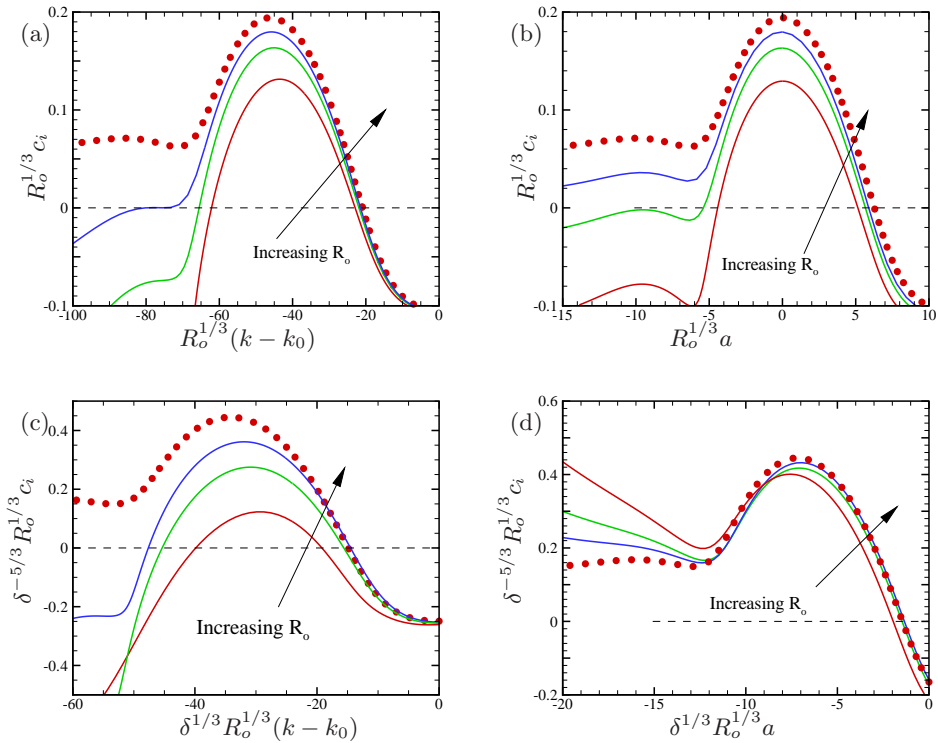
\begin{figure}
 \begin{center}
\includegraphics[width = 0.48\textwidth] {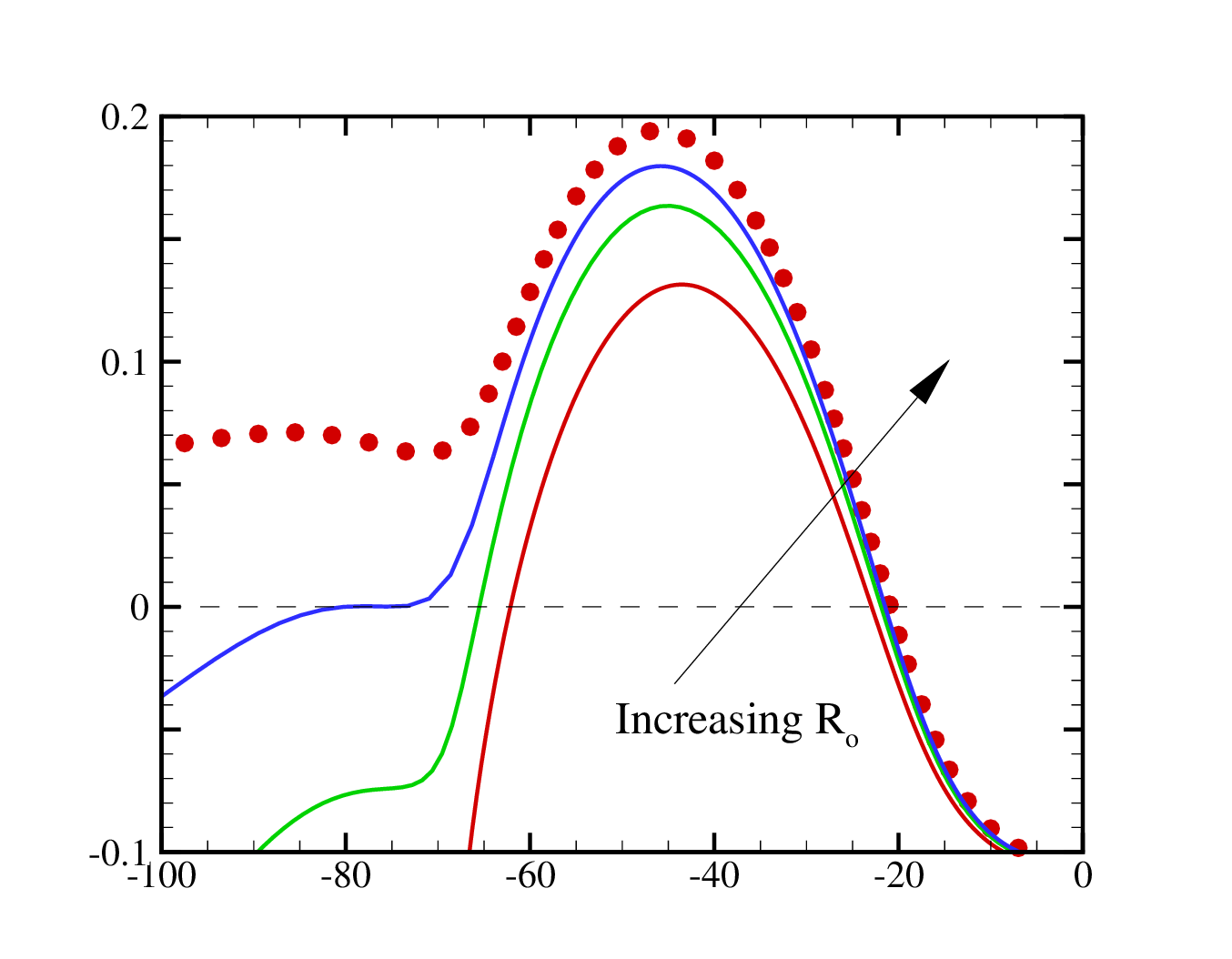}
     \put(-120,1){$R_o^{1/3}(k-k_0)$}\put(-180,55){\rotatebox{90}{$R_o^{1/3}c_{i}$}} \put(-180,115){(a)}
\includegraphics[width = 0.48\textwidth] {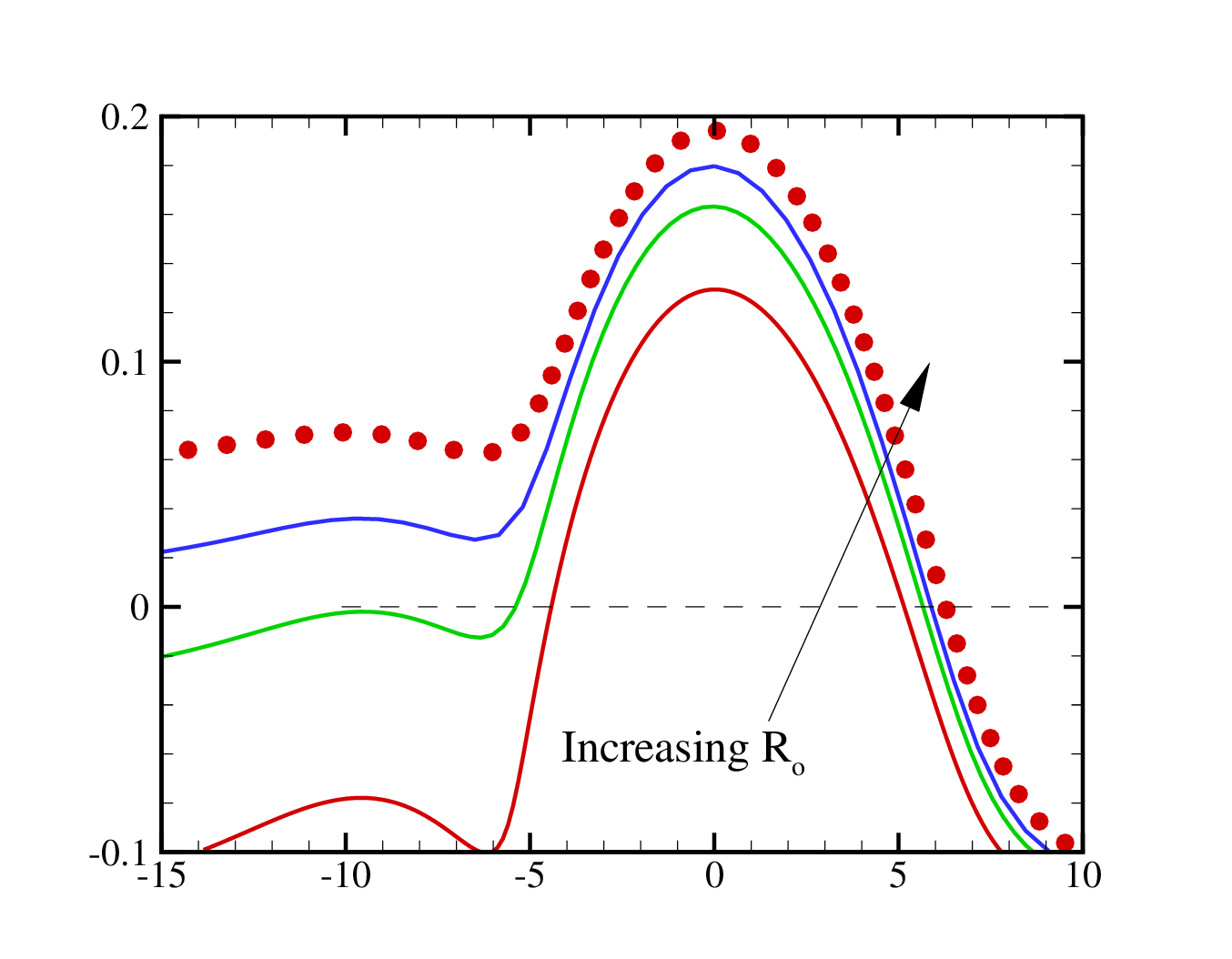}
     \put(-120,1){$R_o^{1/3}a$} \put(-180,55){\rotatebox{90}{$R_o^{1/3}c_{i}$}}\put(-180,115){(b)}
\\
  \includegraphics[width = 0.48\textwidth] {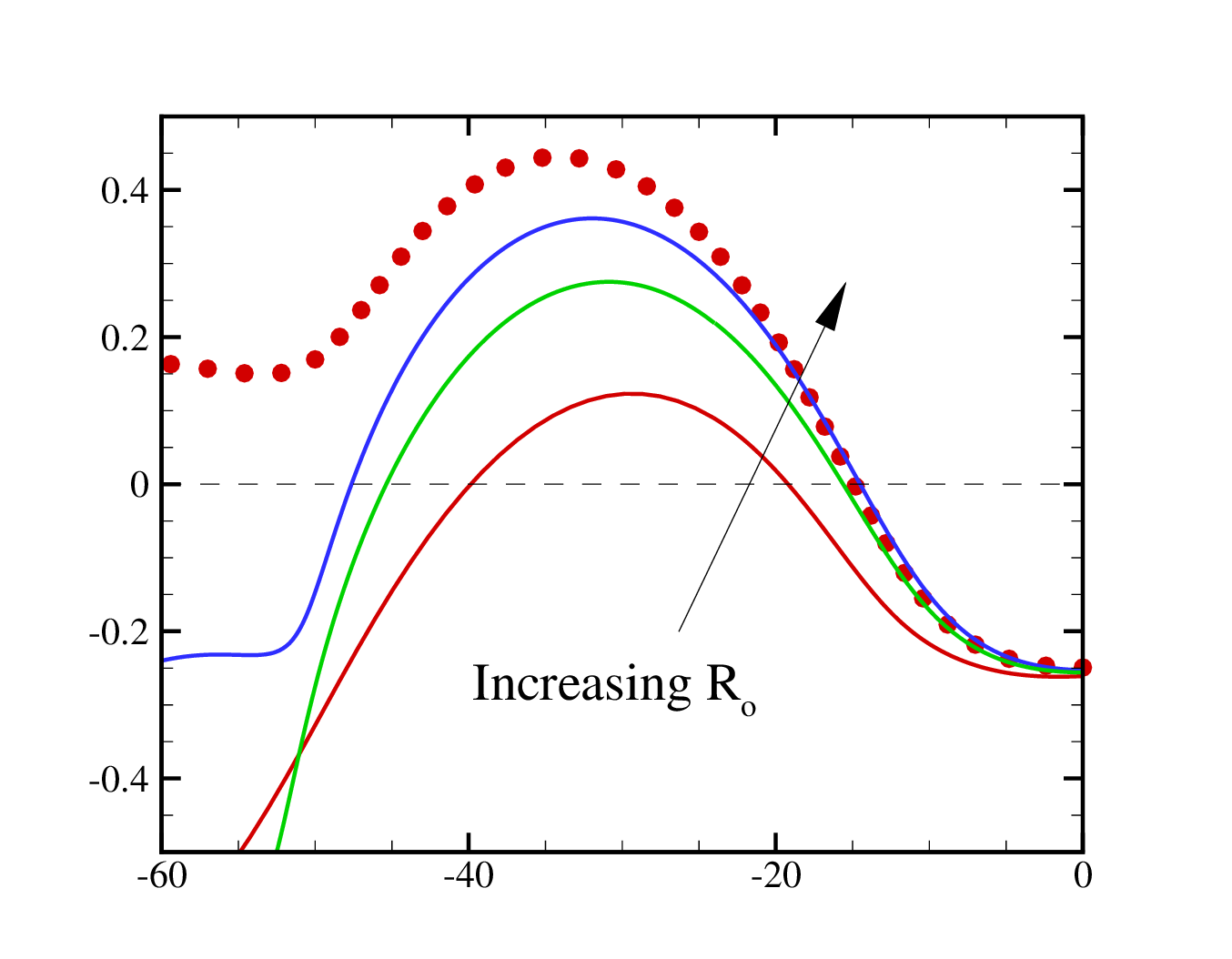}\put(-185,50){\rotatebox{90}{$\delta^{-5/3}R_o^{1/3}c_{i}$}}
     \put(-120,1){$\delta^{1/3}R_o^{1/3}(k-k_0)$}\put(-180,115){(c)}
     \includegraphics[width = 0.48\textwidth] {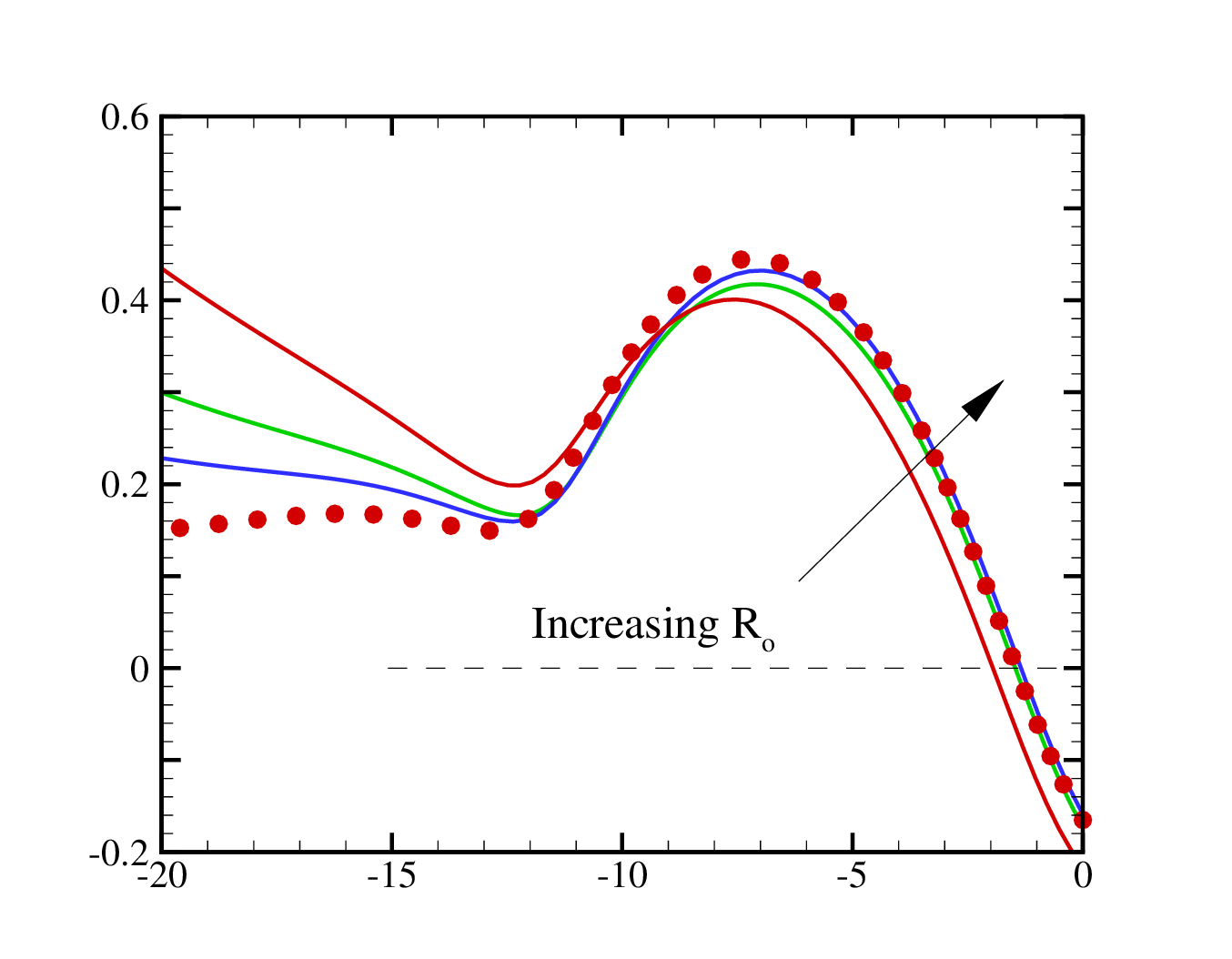}
     \put(-120,1){$\delta^{1/3}R_o^{1/3}a$}\put(-185,50){\rotatebox{90}{$\delta^{-5/3}R_o^{1/3}c_{i}$}}\put(-180,115){(d)}
   \caption{The scaled growth rate. In panels (a) and (b) the wide gap $\eta=5/7$ is used with $m=2$. The red, green and blue curves are the linearised Navier-Stokes results for $R_o= 10^9$, $10^{10}$ and $10^{11}$, respectively. The symbols are the asymptotic results. (a) $a=0$; (b) $k=k_0+R_o^{-1/3}k_1$ with $(k_0,k_1)=(1.651,-46.0)$.
   The symbols are 
   computed by the wide-gap formula (\ref{diswide}). 
%
 Panels (c) and (d) are similar results but for the narrow gap $\eta=0.99$ with $m=0.978/\delta$. (c) $a=0$; (d) $k=k_0+R_o^{-1/3}k_1$ with $(k_0,k_1)=(1.15,-10)$.
   (c) $(k_0,a_1)=(1.442,0)$; (d) $(k_0,k_1)=(1.15,-10)$. The asymptotic computation uses the narrow-gap limit formula (\ref{I1I2gen}) with $\alpha=0.978$. 
   %
}
\label{fig:comparison1}
\end{center}
\end{figure}
Now, let us explore the connection between the numerical analysis presented in this section and the narrow-gap limit analysis discussed in section 2.3.
 Recall that in the latter analysis, we required $Re = \delta R_o = \sqrt{1 - \eta} R_o$ to be large in order to define the small parameter $\overline{\epsilon}=Re^{-1/3}$. Here, the overline is added to distinguish from $\epsilon=R_o^{-1/3}$ defined in section 3.1; likewise, we use $\overline{k}_0, \overline{k}_1,\overline{c}$, and $\overline{c}_1$ to denote the corresponding quantities appeared in section 2.3.
The two wavenumber expansions $k=\overline{k}_0+\overline{\epsilon} \overline{k}_1+\dots$ and $k=k_0+\epsilon k_1+\dots$ imply the equivalence $\overline{k}_0=k_0$ and $\overline{k}_1=\delta^{1/3}k_1$. Of course, in order to observe the convergence to the narrow-gap result, we must set $\alpha=\delta m$ because the wavenumber $\alpha$ is defined on the variable $x$. Also, from the growth rate equality $i\alpha Re \overline{c}=imR_o c$, we expect $\overline{c}_1=\delta^{-5/3}c_1$ to hold. 
Moreover, we write $\overline{a}_1=\delta^{1/3}a_1$ because, with this and the above rescaling, in the limit $\eta\rightarrow 1$, the dispersion relation (\ref{diswide}) reduces to
\begin{eqnarray}
-\frac{1}{ \alpha^{1/3}} \frac{i^{5/3}\text{Ai}'(\overline{\xi}_0)}{ \kappa(\overline{\xi}_0)}+\overline{c}_1 (I_1-1)+\overline{k}_1I_2 +\overline{a}_1(1-\frac{I_1}{2})=0,\label{I1I2gen}
\end{eqnarray}
where $\overline{\xi}_0=i^{7/3}\alpha^{1/3}(\overline{c}_1-\overline{a}_1)$. This is precisely a generalisation of (\ref{I1I2}) for non-zero rotation rates.
The convergence of $k_0$ to the narrow gap result as $\eta\rightarrow 1$ is graphically evident in figure \ref{fig:comparison_leading}b. Figure \ref{fig:comparison2_narrow}b depicts the similar excellent convergence to the narrow-gap result for the $a_1$--$k_1$ relation. The narrow-gap result in this figure was produced by (\ref{I1I2gen}) for $m=0.978/\delta$, so the value of $\overline{k}_1$ at $a_1=0$ corresponds to the lower branch asymptotic curve seen in figure \ref{figck}a.

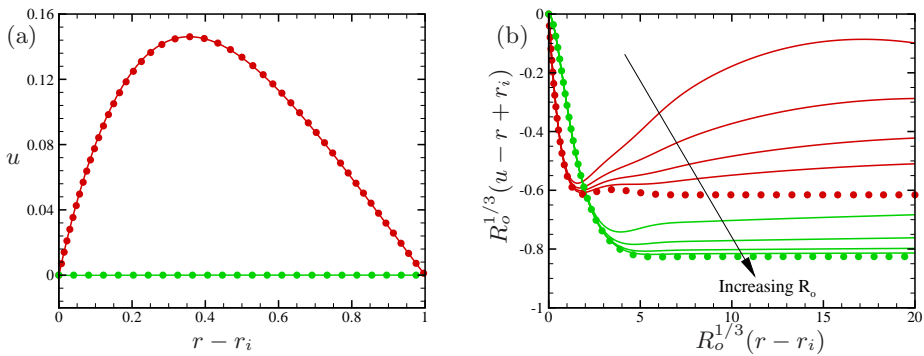
\begin{figure}
 \begin{center}
\includegraphics[width = 0.48\textwidth] {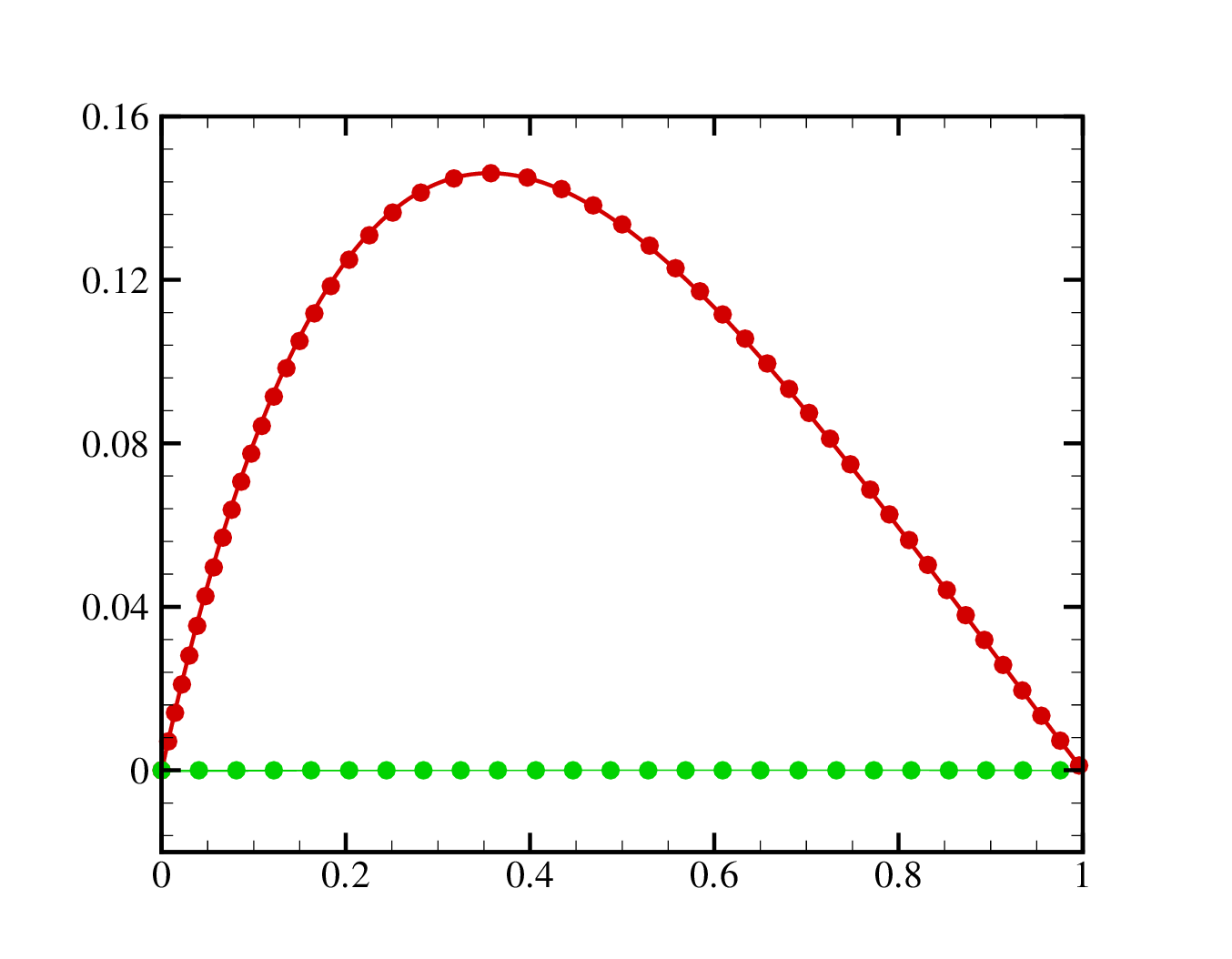}
     \put(-110,1){$r-r_i$}\put(-180,70){$ u$}\put(-180,115){(a)}
\includegraphics[width = 0.48\textwidth] {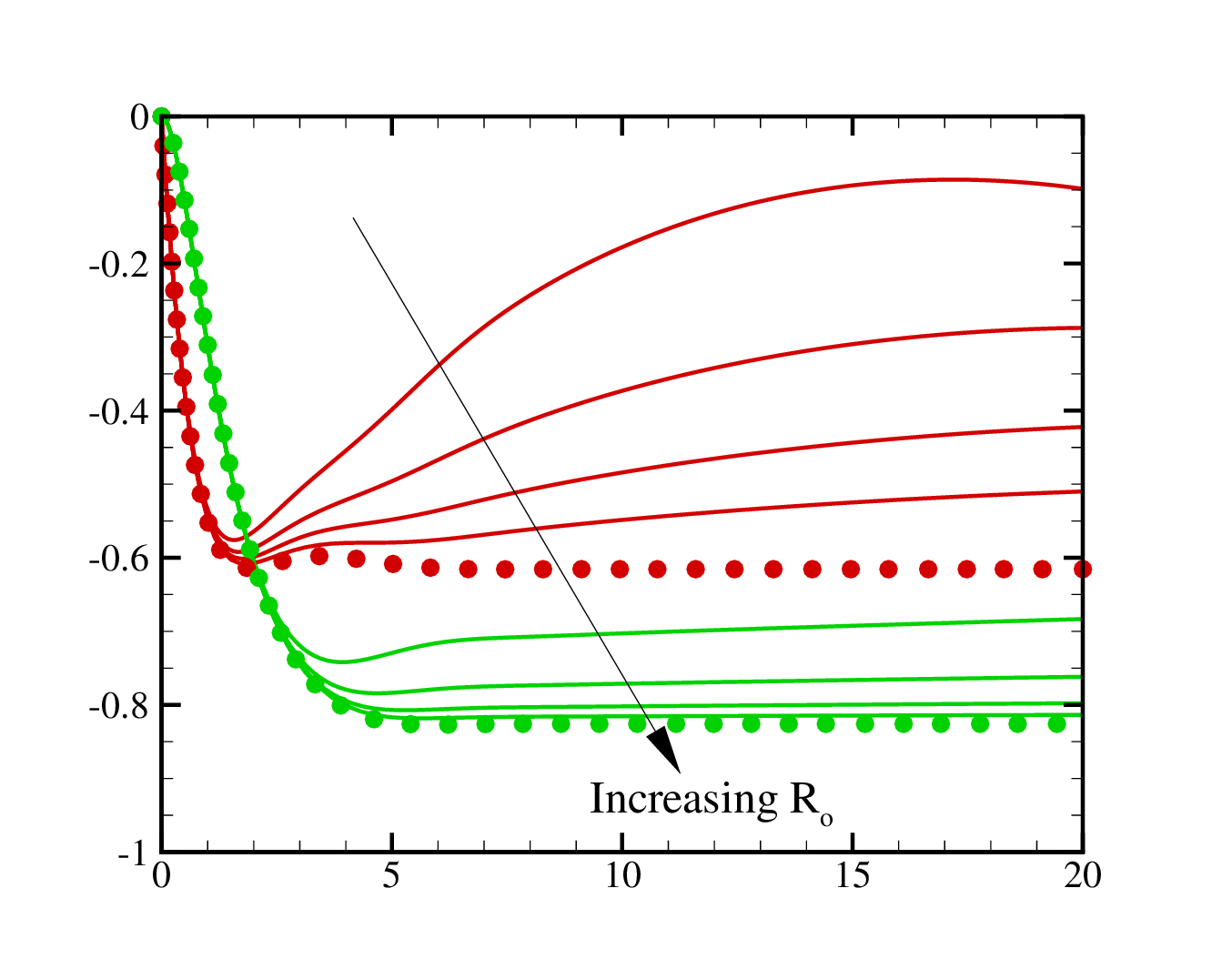}
     \put(-105,1){{$R_o^{1/3}(r-r_i)$}}\put(-185,40){\rotatebox{90}{$R_o^{1/3} (u-r+r_i)$}}\put(-180,115){(b)}
   \caption{Comparison of the eigenfunctions between the leading order asymptotic predictions (symbols) and the linearised Navier-Stokes results (lines) with $R_o=10^{10}, \ 10^{11},\ 10^{12}$ and $10^{13}$. The parameters are $\eta=5/7$, $a=0$, $m=2$ and $k= 1.65$, where the mode is unstable.
      Red: real part; green: imaginary part.
   (a) Comparison in terms of the inviscid core scaling. (b) Comparison in terms of the viscous wall layer scaling.  
}
\label{fig:profile_wide}
\end{center}
\end{figure}
The dispersion relation (\ref{diswide}) can also be used for non-neutral modes. Figure \ref{fig:comparison1} presents the scaled growth rate computed by the asymptotic analysis.
The panels (a) and (b) are the results for $\eta=5/7$, $m=2$, which are used in \cite{Deguchi2017}. The value of $k_0$ is found as 1.651 from (\ref{eq:normal_main}), and the variation of $c_{1i}=\Im(c_1)$ on $k_1$ for $a_1=0$ can be found by (\ref{diswide}) as circles in panel (a). The solid curves represent the full Navier-Stokes computations based on (\ref{eq:OS}), which, as expected, converge to the asymptotic results as $Re$ increases. To compare the results we plotted $R_o^{1/3}c_i$ against $R_o^{1/3}(k-k_0)$; recall the scaling described just above (\ref{expcore1}). The flow becomes unstable when $k_1 \approx R_o^{1/3}(k-k_0)$ is smaller than $-20$, with the growth rate reaching its maximum around $k_1=-46.0$. Fixing $k_1$ at this value and varying $a_1$ results in panel (b). An increase in the rotation rate has a stabilising effect. To observe the convergence in the Navier-Stokes results, we set $k=k_0+R_o^{-1/3}k_1$.

Panels (c) and (d) in figure \ref{fig:comparison1} compare the linearised Navier-Stokes results for $\eta=0.99$ (curves) with the narrow-gap asymptotic results (circles). Again, we choose $\alpha=0.978$ used in section 2 for the asymptotic computation. Thus, $k_0$ is found as $1.442$ by (\ref{MMM}), and in panel (c), where $c_{1i}$ is plotted against $k_1$ for $a_1=0$, the neutral point $k_1\approx -14.9$ recovers the approximation (\ref{klower}). In the Navier-Stokes computations, the choice $\eta=0.99$ implies that the suitable azimuthal wavenumber is $m=0.978/\delta=9.78$. 
The asymptotic convergence of the numerical data $(\delta^{-5/3}R_o^{1/3}c_i,\delta^{1/3}R_o^{1/3}(k-k_0))$ towards $(\overline{c}_{1i},\overline{k}_1)$ computed by the formula (\ref{I1I2gen}) is reasonable, considering that there are now two factors causing errors: the finiteness of $\delta$ and $Re^{-1}$. Even if we choose the closest integer value of $m=10$, the result does not differ significantly. The panel (d) shows the similar comparison but now the rotation rate is varied for fixed $k_1=-10$. Overall, we can observe the similar trend as the wide gap case.
Figure \ref{fig:profile_wide} illustrates a comparison between the eigenfunction of the linearised Navier-Stokes equations (\ref{TCFev}) and the solution of the asymptotic problem.
All the above results unequivocally demonstrate the convergence of the Navier-Stokes computation towards the asymptotic results.



Instability occurs for all values of $\eta \in (0,1)$, if $R_o$ is large enough. This is demonstrated in figure \ref{fig:comparison_leading}, where the case of a stationary inner cylinder case ($a=0$) is shown.
Of particular interest from an application point of view is the extent to which the D17 mode can penetrate into the Rayleigh stable region, i.e. how much we can increase the rotation rate $a$ before the flow completely stabilises. Figure \ref{fig:comparison2_narrow} shows that the lower branch type instability always exists as long as $a$ is $O(R_o^{-1/3})$. The upper branch type survives a bit larger $a$ of $O(R_o^{-1/5})$. When $a$ reaches $O(1)$, the stability is determined by the inviscid problem similar to (\ref{eq:normal_main}) but with a critical layer located away from the walls \citep{Drazin_Reid1981, BiGa05, CaMa07}. 
This type of instability may appear when the wavenumbers are fixed and the Reynolds number is increased.
For the narrow-gap limit case, taking long wavelengths $k=O(Re^{-1})$ and $\alpha=O(Re^{-1})$ leads to another large $Re$ asymptotic limit of instability for $a=O(1)$,
However, this limit, similar to that considered in Cowley \& Smith (1985),  cannot be realised for wide gaps, since $m$ must be an asymptotically small non-zero number. Continuing the neutral curve for general $a$ using (\ref{eq:OS}) is a computationally challenging task, so we do not pursue further investigation.


\section{A boundary layer flow over a convex wall}

In this section, we consider a near wall boundary layer flow influenced by weak wall curvature. Rational analysis of this problem in the high Reynolds number asymptotic limit was established by \cite{Hall1982, Hall1983, Hall1988} for concave wall cases. Our theory developed in the previous sections applies when the wall is convex and the magnitude of the G\"ortler number $G$, as defined by Hall, is sufficiently large. To demonstrate this, here we employ the simplest setup: the asymptotic suction boundary layer \citep{Hocking1982,Milinazzo1985,Fransson2003}.

In this flow configuration, the boundary layer thickness remains unchanged downstream due to suction from the wall. Therefore, the complexity associated with non-parallel effects, such as those present in the Blasius boundary layer, can be avoided.

\subsection{Asymptotic suction boundary layer}
The generalised version of the asymptotic suction boundary layer, incorpolating the effect of wall curvature, can be naturally derived from Taylor-Couette flow by manipulating the boundary conditions as follows:
\begin{equation}\label{NSBCS2}
  \mathbf{v}(r_i,\theta,z)=(-S,R_i,0),\quad \mathbf{v}(r_o,\theta,z)=(-\eta S,R_o,0).
\end{equation}
Here, $S$ represents the non-dimensional suction velocity on the inner cylinder, which introduces a uniform radial crossflow; the same flow configuration was studied in \cite{GDS10}.
The base flow solution can be found as
 $(u,v,w)=(-\frac{r_i}{r} S,R_o r \Omega,0)$, with the function $\Omega(r)$ in the azimuthal component is modified as
\begin{eqnarray}
\Omega=Ar^{-r_iS}+B/r^2,\qquad 
A=\frac{1}{r_o^{1-r_iS}}\frac{1-\eta^2 a}{1-\eta^{2-r_iS}},\qquad B=r_i\frac{\eta a-\eta^{1-r_iS}}{1-\eta^{2-r_iS}}.
\end{eqnarray} 
The associated stability problem is governed by equations (\ref{TCFev}) but the linear operator must include the cross-flow effect as $\mathcal{L}=im (\Omega-c)-R_o^{-1}(\partial_r^2+\frac{r_i S}{r}\partial_r+r^{-1}\partial_r-r^{-2}m^2-k^2)$. 

If we take large values of $S$, the base flow develops a boundary layer near the inner cylinder. For simplicity, we set $a=0$. 
Then it is easy to check that as $S\rightarrow \infty$ the base flow behaves as
\begin{eqnarray}
v\sim \frac{R_or_o}{r}+O(S^{-1}), \qquad \text{if} \qquad r-r_i=O(1),\\
v\sim \frac{R_o}{\eta}(1-e^{-y})+O(S^{-1}),\qquad \text{if} \qquad  y=S(r-r_i)=O(1).
\end{eqnarray}
This observation suggests that the boundary layer thickness is $O(S^{-1})$. 

To clarify the connection with the standard boundary layer analysis, we apply the following manipulations:
\begin{enumerate}
\item Replace the length scale with one based on boundary layer thickness, using $y$ and the wavenumbers $\beta\equiv S^{-1}k$ and $\alpha\equiv m/Sr_i$.
\item Swap $u$ and $v$ to express the equations in the familiar form used the boundary layer analysis.
Also, rewrite $\eta p$ as $p$.
\item Redefine the Reynolds number as $Re\equiv R_o/\eta S$, and write $u_b\equiv r_i\eta \Omega$ and $C=r_i\eta c$.
\end{enumerate}
These transformations recast the stability problem into
\refstepcounter{equation}\label{trans}
$$
 \mathcal{L}u+(2\frac{u_b}{Sr_i} +\frac{r}{r_i} \frac{du_b}{dy})v-\frac{r_i}{r^2 S Re}v+\frac{1}{ S Re}\Big(\frac{u}{Sr^2}-\frac{2i r_i\alpha v}{r^2}\Big)=-i\alpha \frac{r_i}{r} p,\eqno{(\theequation b)}
$$
$$
 \mathcal{L}v-2\frac{u_b}{Sr_i}u +\frac{r_i}{r^2 S Re}v+\frac{1}{ S Re}\Big(\frac{v}{Sr^2} +\frac{2 i r_i\alpha u}{r^2}\Big)=- \frac{dp}{dy},\eqno{(\theequation a)}
$$
$$
 \mathcal{L}w=-i\beta{ p},\eqno{(\theequation c)}$$
$$
i\alpha \frac{r_i}{r}u+\frac{dv}{dy}+\frac{v}{Sr}+i\beta w=0,\eqno{(\theequation d)}
$$
where $\mathcal{L}=i\alpha (u_b-C)-Re^{-1}(\partial_y^2+\frac{r_i }{r}\partial_y+S^{-1}r^{-1}\partial_y-(r_i/r)^2\alpha^2-\beta^2)$. 
If $S$ is large, the stability problem may be simplified to:
\begin{subequations}\label{OSeq_ASBL}
\begin{eqnarray}
\mathcal{L}{u}+u_b'{v}=-i{\alpha} {p},\qquad
\mathcal{L}{v}=-{p}'+g u_b{u},\\
\mathcal{L}{w}=-i{\beta}{p},\qquad
i{\alpha} {u}+{v}'+i{\beta}\,{w}=0,
\end{eqnarray}
\end{subequations}
where $\mathcal{L}=\ri \alpha (u_b-c)-Re^{-1}(\partial_{y}^2+\partial_{y}-\alpha^2- \beta^2)$, $u_b=1-e^{-y}$, and $g=2/Sr_i$. The primes denote derivatives with respect to $y$. At $y=0$ the no-slip conditions $u=v=w=0$ are imposed, and we assume the perturbations vanish as $y\rightarrow \infty$.

The constant $g$ is the parameter that represents the ratio of the boundary layer thickness to the radius of curvature of the wall. 
When the term proportional to $g$ is absent, the above system reduces to the usual Orr-Sommerfeld and Squire equations for the asymptotic suction boundary layer, where the onset of the instability is well-known to occur at $(\alpha,\beta,Re)=(0.15547,0,27189)$ \cite{Hocking1982,Drazin_Reid1981}.
The mechanism of this instability is essentially of the TS wave type.
Note also that G\"ortler vortex type instabilities appear when $G=gRe^2$ is an $O(1)$ negative number, akin to Hall's work on growing boundary layers. In this case, we need to recover several terms we omitted from (\ref{trans}). However, those terms do not affect our analysis that follows. 
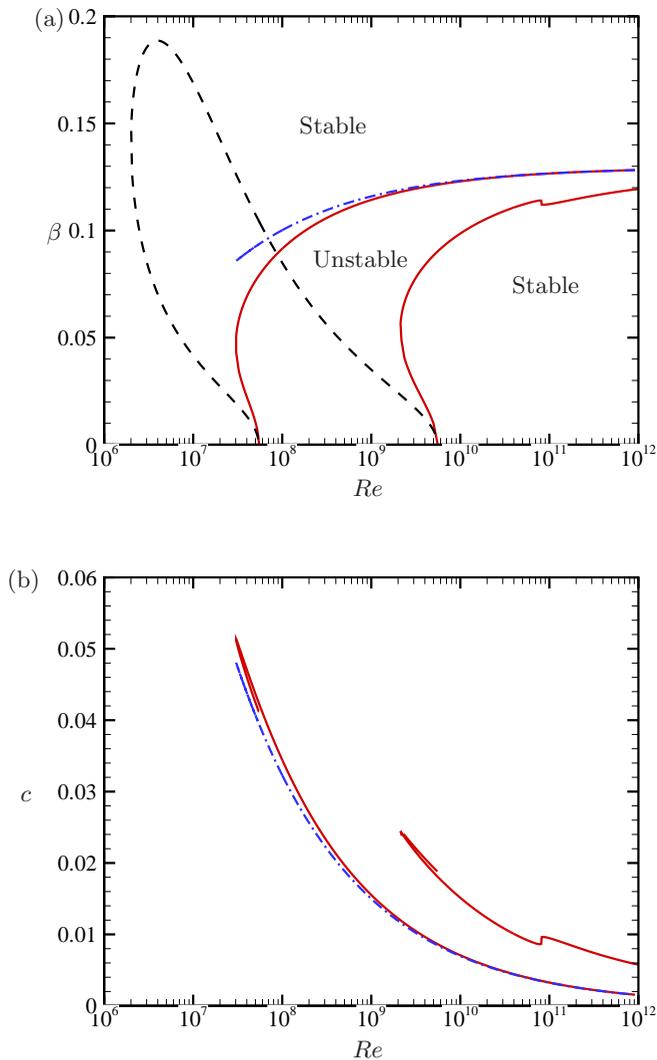
\begin{figure}
 \begin{center}
           \includegraphics[width = 0.7\textwidth] {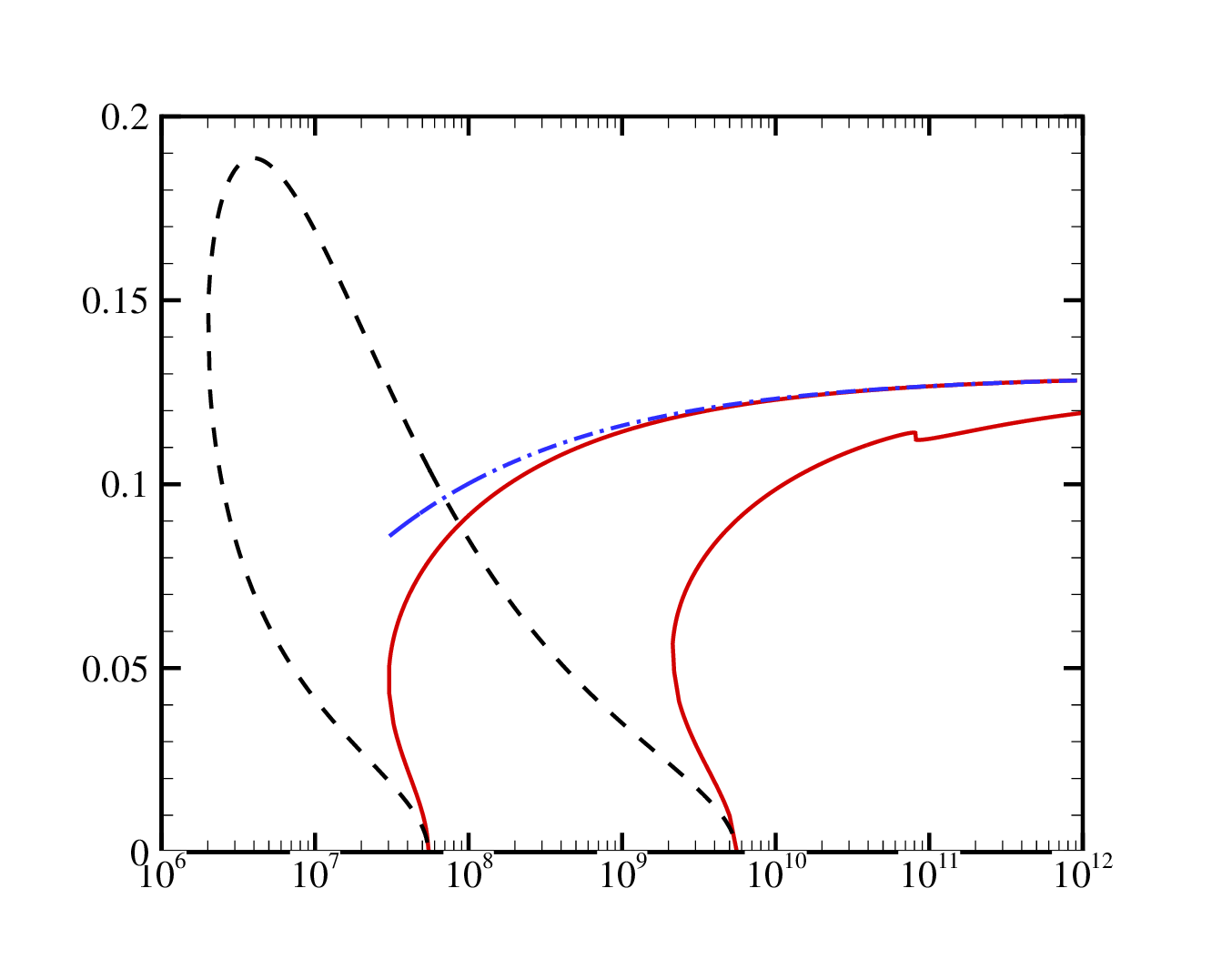}
           \put(-260,180){(a)}
           \put(-155,90){Unstable}
           \put(-80,80){Stable}
    \put(-160,140){Stable} \put(-140,4)
    {$Re$}\put(-255,100){\rotatebox{0}{$\beta$}}\\
               \includegraphics[width = 0.7\textwidth] {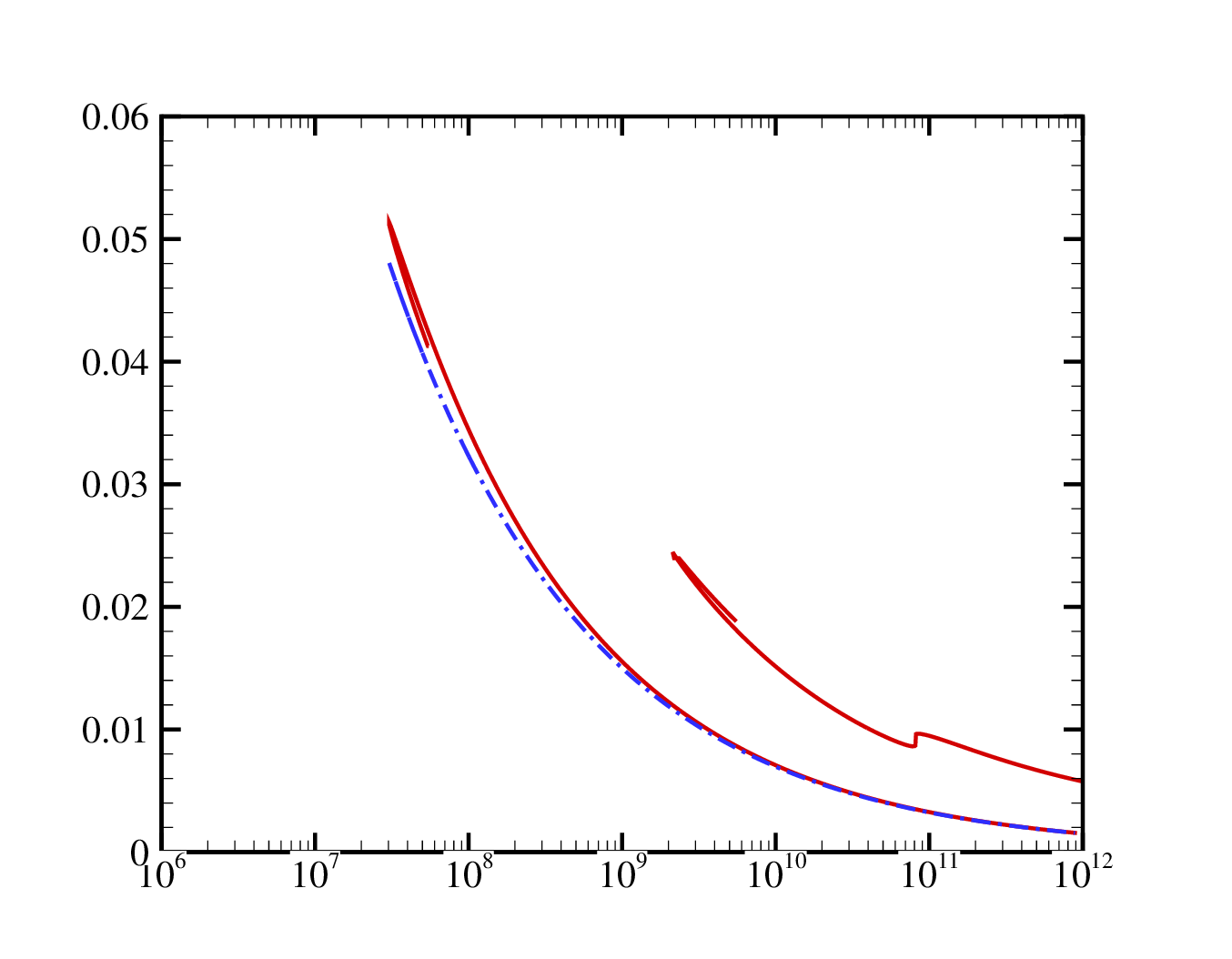}
           \put(-140,4)
    {$Re$}             \put(-270,180){(b)}
    \put(-265,100){\rotatebox{0}{$c$}}
             \caption{
             Linear stability of the asymptotic suction boundary layer computed by (\ref{OSeq_ASBL}). The streamwise wavenumber is fixed at $\alpha=0.0036$.
             (a) 
              The red solid curves represent the neutral curves for the convex wall case $g=2\times 10^{-4}$ (with stable/unstable regions indicated by the labels), while the black dashed curve corresponds to the flat wall case $g=0$. The blue dot-dashed curve is the asymptotic result given by (\ref{ASBLasym}) for  $g=2\times 10^{-4}$. (b) The phase speed of the neutral modes for $g=2\times 10^{-4}$. 
             }
\label{fig:ASBL_OS}
\end{center}
\end{figure}

Figure \ref{fig:ASBL_OS} presents the numerical results of the eigenvalue problem (\ref{OSeq_ASBL}). The streamwise wavenumber is fixed at $\alpha=0.0036$ for simplicity, and the figure plots the neutral Reynolds numbers against the spanwise wavenumber $\beta$. The two red curves correspond to the case $g=2\times 10^{-4}$, where the flow becomes unstable in the region between them. For comparison, the black dashed curve indicates the result for $g=0$. 
The figure demonstrates that even small variations in $g$ can cause significant changes in the shape of the neutral curve. 
Comparing the black and red curves reveals that the effect of wall curvature generates a new `D17' instability in a narrow band of $\beta$ at large Reynolds number regime. The asymptotic behavior of the D17 modes can be explained by the lower and upper branch theories, analogues to those discussed in section 2, with the lower branch elaborated in section 4.2.

For spanwise-independent perturbations, we can theoretically show that the stability remains unchanged from the $g=0$ case. Therefore, the unstable region at $g=2\times 10^{-4}$ represents a mixture of the TS mode and the D17 mode. Varying $\beta$ from zero along the red curve illustrates a smooth transition from TS waves to D17 modes, with the Reynolds number reaches the minimum value $Re=3.01\times 10^7$ at $\beta=0.0467$. 
Note that this Reynolds number is not a critical value, as it is not optimised across all wavenumber pairs. Clearly, for any value of $g$, the critical Reynolds number cannot fall below 27819, which occurs at $\beta=0$. We have confirmed that the critical Reynolds number remains at this value for $g=2\times 10^{-4}$. In general, for relatively large $\alpha$, small values of $g$ have no significant impact on stability. Numerical computations were necessary to verify this, as Squire's theorem does not hold for $g\neq 0$. 

\subsection{Asymptotic analysis}
Here, we conduct a lower branch type matched asymptotic expansion analysis for the stability problem (\ref{OSeq_ASBL}). The key assumptions in the asymptotic analysis are $Re\gg 1$ and $g\gg Re^{-2}$. The second condition is satisfied at the high-Reynolds number limit if $g>0$ is held constant.

The analysis proceeds along the same lines as in section 2.3. 
An inviscid main layer emerges at which $y=O(1)$, while the no-slip conditions are satisfied via the viscous wall layer of thickness $O(\epsilon)$ with $\epsilon=(gRe^2)^{-1/6}\ll 1$. Those regions are somewhat similar to the main deck and lower deck of the triple-deck theory, respectively, but a structure like the upper deck does not emerge. Using the small parameter $\epsilon$, we expand the spanwise wavenumber and the phase speed as $\beta=\beta_0+\epsilon \beta_1+\cdots$ and $C=\epsilon C_1+\cdots$, respectively. Also, we rescale the streamwise wavenumber as $\alpha=g^{1/2}\alpha_0$. 

In the main layer, we write
\begin{eqnarray}\label{GoMainEx}
\left [
\begin{array}{c}
u\\
v\\
w\\
p
\end{array}
\right ]
=
\left [
\begin{array}{c}
u_0(y)+\epsilon u_1(y)+\cdots\\
g^{1/2}(v_0(y)+\epsilon v_1(y)+\cdots)\\
g^{1/2}(w_0(y)+\epsilon w_1(y)+\cdots)\\
g(p_0(y)+\epsilon p_1(y)+\cdots)
\end{array}
\right ].
\end{eqnarray}
Substituting (\ref{GoMainEx}) into (\ref{OSeq_ASBL}), the leading order equations can be combined to yield
\begin{eqnarray}\label{Golead}
\mathscr{L}v_0=0,\qquad \mathscr{L}=\partial_y^2-\beta_0^2 -\frac{u_b''}{u_b} +\frac{\beta_0^2}{\alpha_0^2}\frac{u_b'}{u_b}.
\end{eqnarray}
While, as expected, the next order equations yield the inhomogeneous equation
\begin{eqnarray}\label{Gonext}
\mathscr{L}v_1=\frac{C_1}{u_b}\left (v_0''-\beta_0^2v_0-\frac{\beta_0^2}{\alpha_0^2}\frac{u_b'}{u_b}v_0\right )+2\beta_0\beta_1\left (1-\frac{1}{\alpha_0^2}\frac{u_b'}{u_b}\right )v_0.
\end{eqnarray}
The solution of (\ref{Golead}) satisfying $v_0=0$ at $y=0$ and $v_0\rightarrow 0$ as $y\rightarrow \infty$ can be found as
\begin{eqnarray}
v_0=\lambda^{-1}u_b e^{-\beta_0y}\label{Gov0}
\end{eqnarray}
only when 
\begin{eqnarray}\label{ASBLab}
\beta_0=2\alpha_0^2. 
\end{eqnarray}
The constant $\lambda$ represents the value of $u_b'$ at $y=0$, and the factor $\lambda^{-1}$ in (\ref{Gov0}) normalises $v_0'=1$ at $y=0$. Note that when the sign of $g$ is negative (i.e., in the case of a concave wall), constructing a non-trivial solution becomes impossible.

The solvability condition of the inhomogeneous equation (\ref{Gonext}) then determines the dispersion relation. The far-field condition $v_1\rightarrow 0$ as $y\rightarrow \infty$ remains unchanged from the leading order problem. 
Near the wall, however, the effect of the displacement velocity from the viscous wall layer must be accounted, such that $v_1=V_d$ at $y=0$. The constant $V_d$ can be found by the wall layer analysis.

The viscous wall layer expansions are
\begin{eqnarray}
\left [
\begin{array}{c}
u\\
v\\
w\\
p
\end{array}
\right ]
=
\left [
\begin{array}{c}
U_0(Y)+\cdots\\
g^{1/2}(\epsilon V_0(Y)+\cdots)\\
g^{1/2}(W_0(Y)+\cdots)\\
g(\epsilon P_0(Y)+\cdots)
\end{array}
\right ].
\end{eqnarray}
The no-slip conditions are $U_0=V_0=V'_0=0$ at $Y=0$.
The leading order wall-normal component $V_0$ satisfies 
\begin{eqnarray}
(i\alpha  (\lambda Y-C_1)-\partial_Y^2)V_0''=0,
\end{eqnarray}
which is the same equation as (\ref{wallU}). Therefore, the displacement velocity $V_d=\lim_{Y\rightarrow \infty} V_0-Y$ is obtained as
\begin{eqnarray}\label{VdGo}
V_d= -\frac{i^{5/3}\text{Ai}'(\xi_0)}{(\lambda \alpha_0)^{1/3} \kappa(\xi_0)}-\frac{C_1}{\lambda},\qquad \xi_0=\frac{i^{7/3}(\lambda \alpha_0)^{1/3}C_1}{\lambda}.
\end{eqnarray}

Multiplying (\ref{Gov0}) to the inhomogeneous equation (\ref{Gonext}) and applying integration by parts, we obtain a dispersion relation similar to (\ref{I1I2}):
\begin{eqnarray}\label{dispASBL}
-\frac{1}{(\lambda \alpha_0)^{1/3}} \frac{i^{5/3}\text{Ai}'(\xi_0)}{ \kappa(\xi_0)}+{C_1} (I_1-\frac1{{\lambda}})+\beta_1I_2 =0,
\end{eqnarray}
where
\begin{eqnarray}
I_1=\int_0^{\infty}\frac{4\beta_0u_b'-u_b''}{\lambda^2}e^{-2\beta_0y}dy,\qquad
I_2=\int_0^{\infty}\frac{4u_bu_b'-2\beta_0u_b^2}{\lambda^2}e^{-2\beta_0y}dy,
\end{eqnarray}
and $\xi_0=i^{7/3}(\lambda \alpha_0)^{1/3} C_1/{{\lambda}}$. For the neutral disturbances, the result simplifies to
\begin{eqnarray}\label{C1B1}
C_1=\frac{\lambda s_0}{(\lambda \alpha_0)^{1/3}},\qquad 
\beta_1 =\frac{q_0+s_0 (1-\lambda I_1)}{(\lambda \alpha_0)^{1/3}I_2 } ,
\end{eqnarray}
where $q_0$ is the number defined in (\ref{k1int}), and recall that $s_0\approx 2.2972$.

\begin{figure}
 \begin{center}
           \includegraphics[width = 0.48\textwidth] {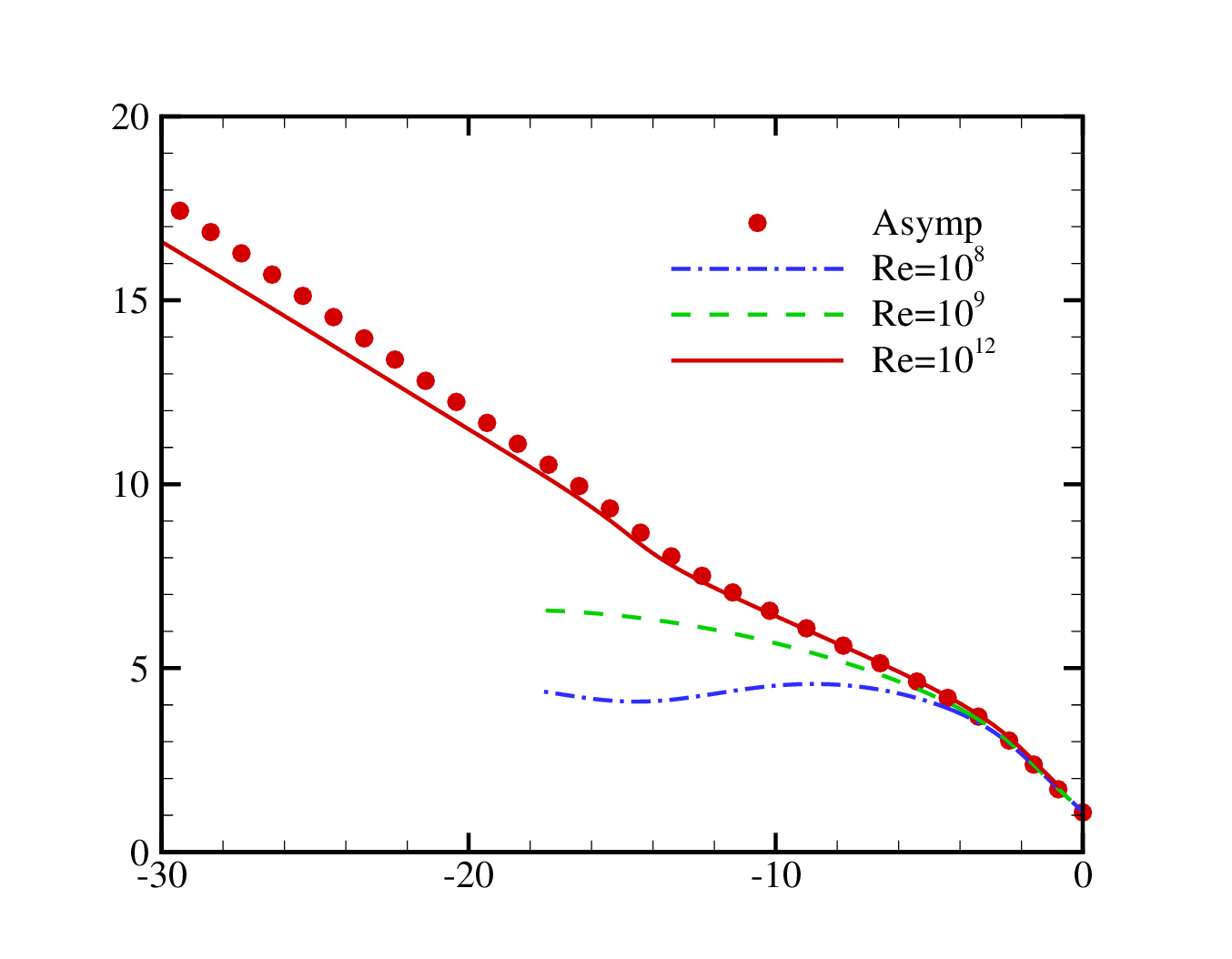}
     \put(-120,0){$( \beta-\beta_0)/\epsilon$}\put(-185,55){\rotatebox{90}{$ C_r/\epsilon$}}\put(-185,120){(a)}
     \includegraphics[width = 0.48\textwidth] {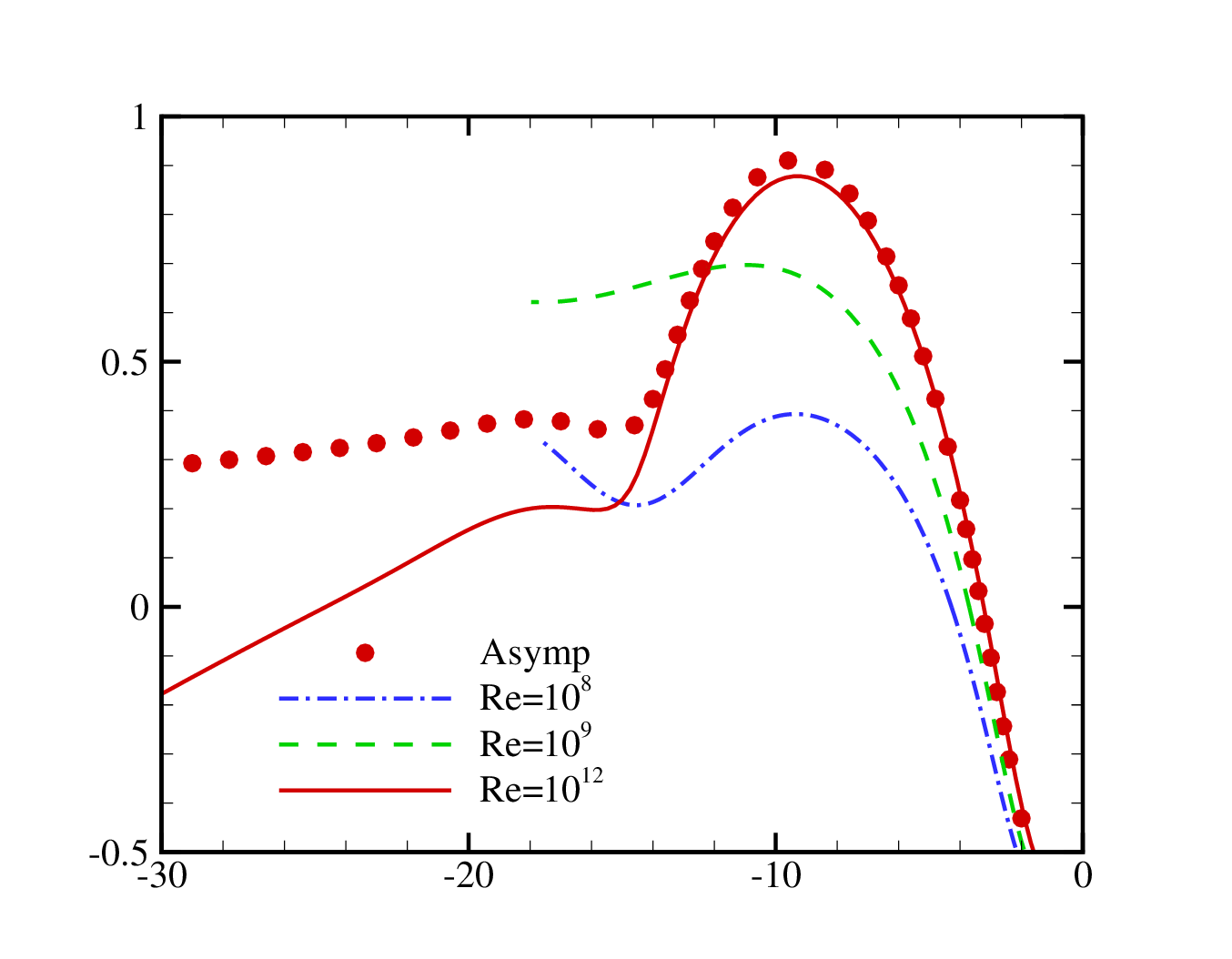}
     \put(-120,0){$( \beta-\beta_0)/\epsilon$}\put(-185,55){\rotatebox{90}{$C_i/\epsilon$}}\put(-185,120){(b)}
        \caption{
Asymptotic convergence of the phase speed for $\alpha=0.0036$, $g=2\times 10^{-4}$.
The curves are computed by (\ref{OSeq_ASBL}). The points are the asymptotic result (\ref{dispASBL}).
(a) Real part; (b) imaginary part.        
        }
\label{fig:BL_ASYMP}
\end{center}
\end{figure}

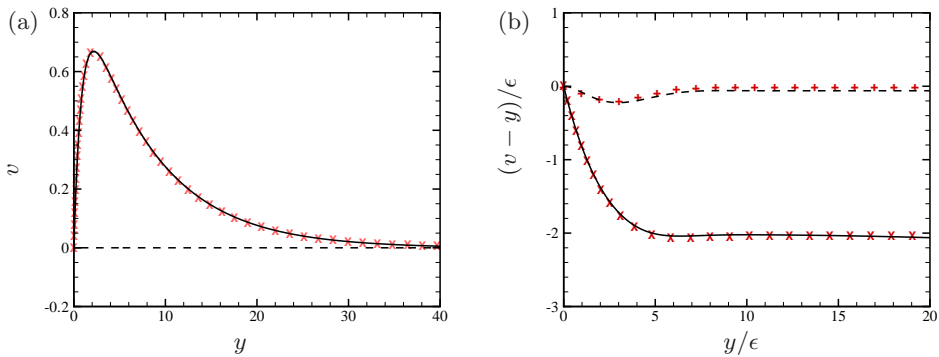
\begin{figure}
 \begin{center}
           \includegraphics[width = 0.48\textwidth] {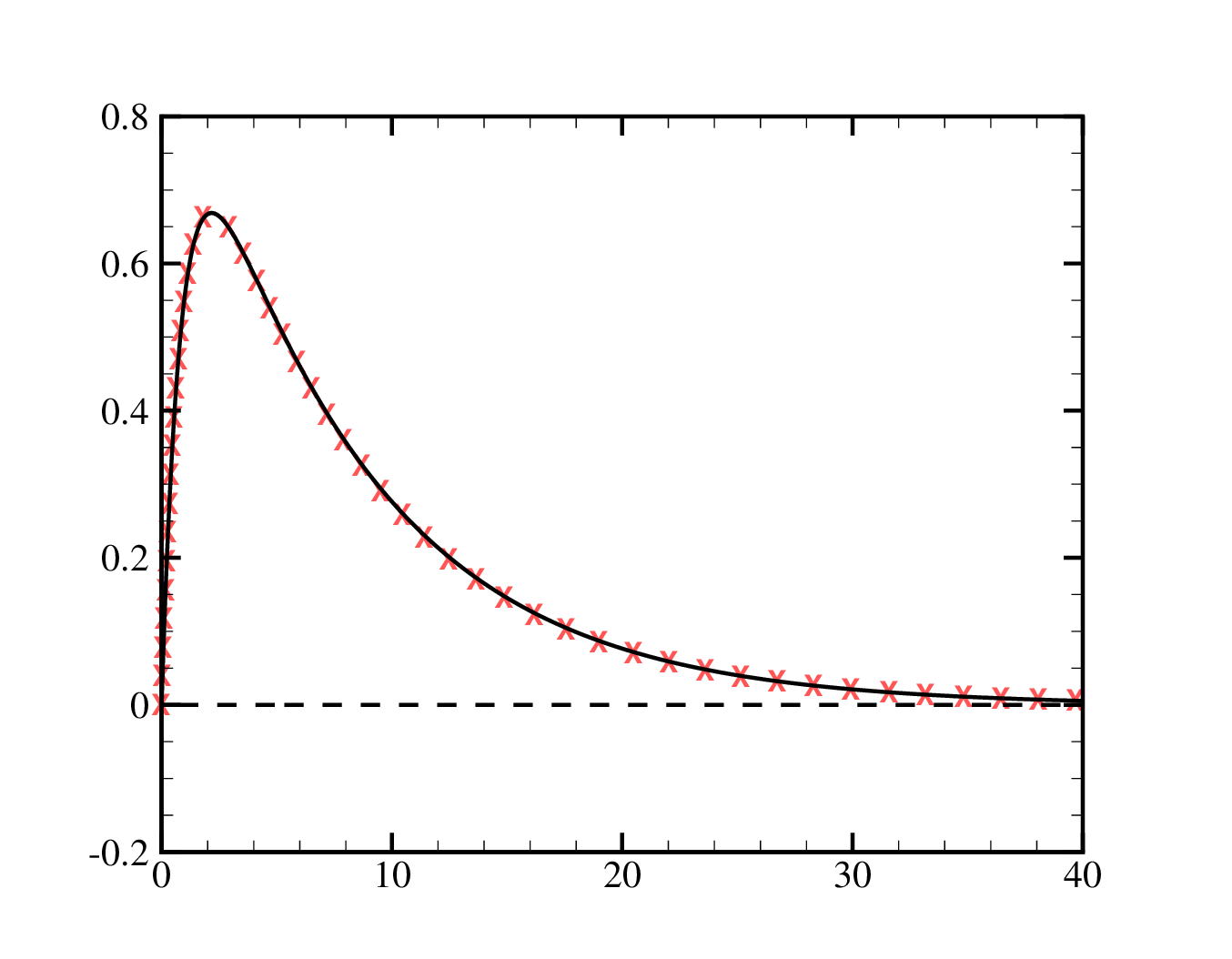}
     \put(-100,0){$y$}\put(-185,65){\rotatebox{90}{$v$}}\put(-185,120){(a)}
     \includegraphics[width = 0.48\textwidth] {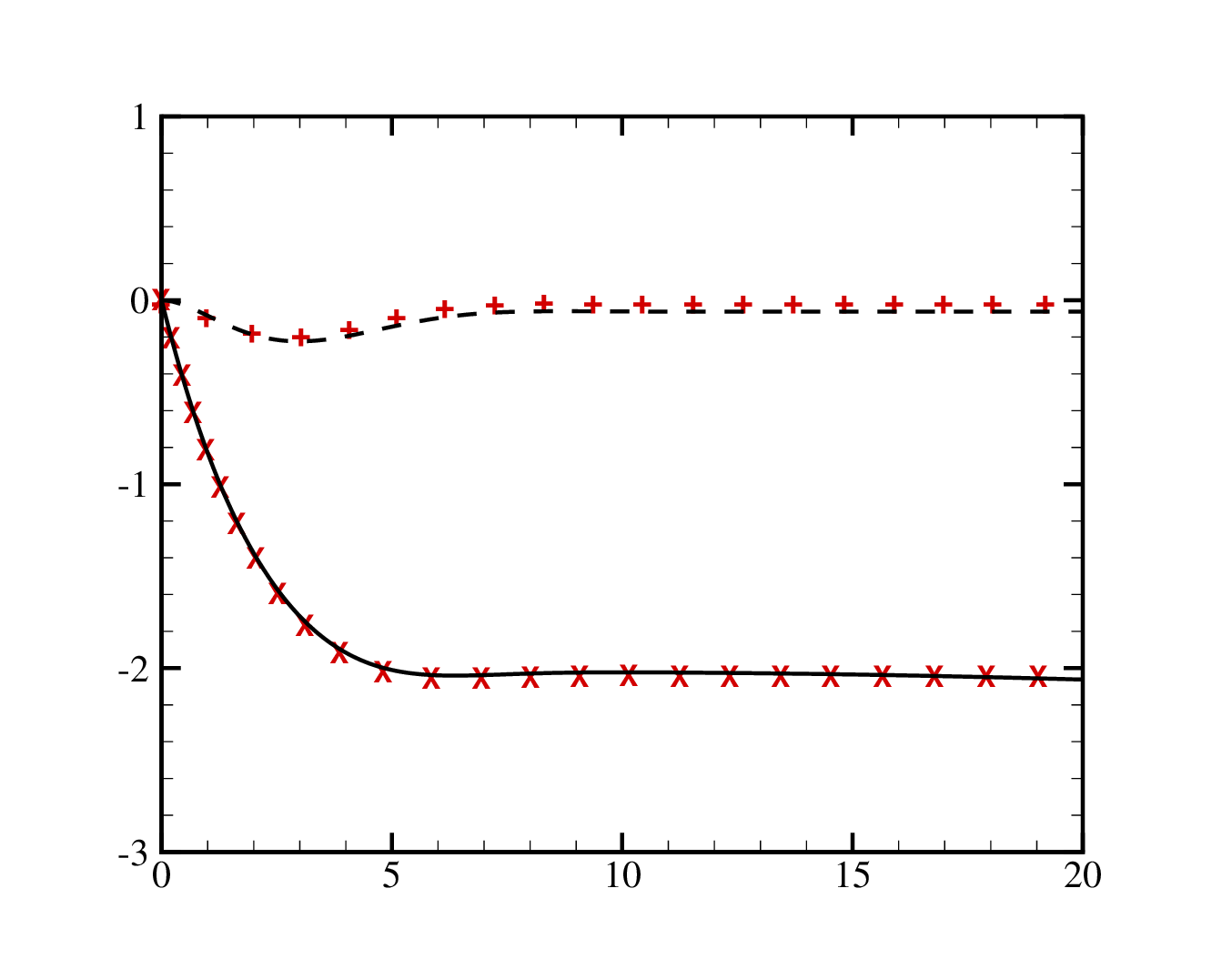}
     \put(-100,0){$y/\epsilon$}\put(-185,65){\rotatebox{90}{$(v-y)/\epsilon$}}\put(-185,120){(b)}
        \caption{Comparison of the eigenfunctions at the lower-branch neutral point at $(Re,g,\alpha,\beta)=(10^{12},2\times 10^{-4},0.0036,0.130-3.30\epsilon)$. The solid and dashed curves are the real and imaginary parts of $v$ computed by (\ref{OSeq_ASBL}). The symbols are the corresponding asymptotic predictions. 
        (a) Comparison in the core scaling. The symbols are the leading order solution $v_0$ found in (\ref{Gov0}). (b) comparison in the wall-layer scaling. The symbols are $V_0 - Y$, where $V_0$ is obtained in the same manner as in the derivation of (\ref{U0Airy}).
        }
\label{fig:ASBL_profile}
\end{center}
\end{figure}
So far, we considered a general $u_b(y)$, allowing the results to be applicable to other boundary layer flows. The specific outcomes for the asymptotic suction boundary layer are obtained by substituting $u_b=1-e^{-y}$ and $\lambda=1$. 
For example, the integrals in (\ref{C1B1}) can be solved analytically, resulting in
\begin{eqnarray}\label{ASBLCB}
C_1=\frac{s_0}{\alpha_0^{1/3}},\qquad 
\beta_1 =\frac{8\alpha_0^4+6\alpha_0^2+1}{\alpha_0^{1/3} } \left ( q_0
- \frac{4s_0\alpha_0^2}{4\alpha_0^2+1} \right ).
\end{eqnarray} 
For the parameters used in figure \ref{fig:ASBL_OS}, the leading order wavenumbers are obtained as $(\alpha_0,\beta_0)=(0.255,0.130)$, and we have the approximations
\begin{eqnarray}\label{ASBLasym}
\beta=0.130-3.31 g^{-1/6} Re^{-1/3},\qquad C=3.62 g^{-1/6}Re^{-1/3},
\end{eqnarray}
plotted by the blue dot-dashed curves. When the Reynolds number is large, these analytical results provide an excellent approximation for the `lower branch', which appears as the upper branch in figure \ref{fig:ASBL_OS}a.
Figure \ref{fig:BL_ASYMP} examines the complex phase velocity $C=C_r+iC_i$ near this branch. As noted in section 3, predicting the asymptotic behaviour of the non-neutral modes requires using the full dispersion relation (\ref{dispASBL}). 
As the Reynolds number increases, the numerical data $((\beta-\beta_0)/\epsilon,C/\epsilon)$ converges to the asymptotic result $(\beta_1,C_1)$.
The points where $C_i$ becomes zero correspond to (\ref{ASBLasym}). Figure \ref{fig:ASBL_profile} compares the eigenfunctions at those neutral points. 
Once again, we observe clear convergence of the numerical results to the leading-order asymptotic solution.

\section{Conclusion}

By employing matched asymptotic expansion analysis at high Reynolds numbers, we investigated the physical mechanism underlying the D17 mode instability, explaining why it displays distinct properties compared to classical centrifugal instabilities. Our asymptotic predictions closely match numerical results across different flow configurations at finite Reynolds numbers, demonstrating the robustness of our theoretical approach.

Taylor-Couette flow has played a central role throughout the paper, providing a canonical framework for studying the instability mode. In section 2, we examine the narrow gap limit, assuming the inner cylinder to be stationary, which leads to a system that is neutrally stable according to Rayleigh's stability criterion. We conducted an asymptotic analysis for both the upper and lower branches of the neutral curve. Notably, the structure of the lower branch theory is relatively straightforward and applies across a wide range of Reynolds numbers.
In section 3, we expanded the lower branch theory to study full Taylor-Couette problem. The D17 mode instabilities occur in both Rayleigh-stable and Rayleigh-unstable scenarios, and this finding is consistent across all radius ratios. In section 4, we introduced a radial crossflow to the Taylor-Couette flow, forming an asymptotic suction boundary layer near the inner cylinder. This flow configuration serves as a simple prototype for boundary layer flow over a convex wall. 
The curvature of the wall results in the most unstable mode identified in high-Reynolds number computations being oblique, which is consistent with the asymptotic analysis.


Our numerical results indicate that the D17 mode instability exists only within a very narrow wavenumber range. This phenomenon, elucidated through asymptotic analysis,  likely explains why this mode has remained undetected for so long.
The first step in our theoretical study involves determining the inviscid core structure, which reveals that, at leading order, nontrivial modes can only arise around a specific curve in the wavenumber plane. While the unstable region forms around this curve, its width is asymptotically narrow. The next-order analysis shows that this region is bounded by what we refer to as the upper and lower branches of the neutral curve, a terminology derived from the analysis of TS waves (\cite{Tollmien29}, \cite{Schlichting33}, and \cite{Lin45,Lin1955}). 

To derive the dispersion relation, it is necessary to analyse both the inviscid core and the viscous wall layer near the inner cylinder. The boundary layer structure is similar to, but not identical to, that of TS waves. The D17 mode cannot be identified by considering inviscid theory alone because the instability requires the presence of a viscous boundary layer. 
However, the viscous wall layer theory alone is insufficient, as the three-dimensional characteristics of the D17 mode are dictated by the inviscid core structure.

{
The theoretical framework developed here can be extended to more general boundary layers over a convex wall, including those exhibiting spatial growth. For small wall curvature, the properties of the instability at low Reynolds numbers may be indistinguishable from the TS wave in the flat wall case. However, at sufficiently high Reynolds numbers, the D17 mode, characterised by low frequency, short spanwise wavelength, low phase speed, and an oblique nature, emerges. In boundary layer transition, it is known that the turbulent transition process depends on external factors.
The discovery that even weak wall convexity influences the characteristics of unstable modes may be important for controlling laminar flow over an airplane wing, as, depending on the flow conditions, the D17 mode should be accounted for. 
Furthermore, the fact that the streamwise length scale of the D17 mode is long but remains shorter than the non-parallel spatial development scale of the base boundary-layer profile is interesting from the perspective of receptivity theory. The D17 mode has the potential to interact with freestream acoustic or vortical waves and extended surface imperfections, such as roughness, similar to the high Reynolds number asymptotic theories by \cite{Ruban1984} and \cite{Goldstein1985}.}

Finally, we note that this paper primarily focused on cases where the flow is nearly marginally stable according to Rayleigh's criterion. While this assumption simplified the theoretical discussion, the precise range of parameters where the D17 mode occurs remains undetermined. 
Identifying this range is important for the experimental detection of the D17 mode, which has yet to be observed. The subcritical bifurcation of nonlinear travelling waves reported by \cite{AyDe20} complicates addressing this issue. An asymptotic approach could facilitate the parameter study of nonlinear states. We speculate that the formulation of the theory may require nonlinear critical layer analysis, similar to that developed by \cite{CaMa07}.

\vspace{.4cm}
\backsection[Acknowledgements]{
This research was supported by the Australian Research Council Discovery Project DP230102188. MD is also supported by National Natural Science Foundation of China (grant nos. 92371104, 11988102) and CAS project for Young Scientists in Basic Research (YSBR-087).
}

\backsection[Declaration of Interests]{
The authors report no conflict of interest.
}



\appendix

\section{Operators and functions in the asymptotic analysis of wide gap Taylor-Couette flow}

The operators and functions in (\ref{eq:normal_main}) and (\ref{eq:u_1_wide}) are defined as
\refstepcounter{equation}
$$
\begin{array}{c}\displaystyle
\mathscr{L}=\partial_{r}^2+\Big(\frac{3m^2+k_0^2r^2}{r(m^2+k_0^2r^2)}\Big){\partial_r}+\Big(-\frac{m^2}{r^2}-k_0^2+\frac{1}{r^2}+
\frac{6k^2m^2+4k_0^4r^2}{m^2(m^2+k_0^2r^2)} \vspace{.2cm}\\ \displaystyle
+
\frac{(-3m^4+k_0^2m^2r^2+2k_0^4r^4)\Omega'_0}{m^2r(m^2+k_0^2r^2)\Omega_0}
-\frac{\Omega''_0}{\Omega_0}\Big),
\end{array}\eqno{(\theequation a)}\label{eq:operator_wide}
$$
$$
F_0=\Big(\frac{4k_0^2r(3m^2+2k_0^2r^2)\Omega_0+(-3m^4+3k_0^2m^2r^2+4k_0^4r^4)\Omega'_0}{m^2r(m^2+k_0^2r^2)\Omega^2_0}
-\frac{\Omega''_0}{\Omega^2_0}\Big),\eqno{(\theequation b)}
$$
$$
F_1=2k_0\Big(\frac{4}{m^2}-1+\frac{2m^2}{(m^2+k_0^2r^2)^2}+\frac{2r[m^4+(m^2+k_0^2r^2)^2]\Omega'_0}
{m^2(m^2+k_0^2r^2)^2\Omega_0}\Big),\eqno{(\theequation c)}
$$
$$
F_2=\frac{4k_0m^2r}{(m^2+k_0^2r^2)^2},\eqno{(\theequation d)}
$$
$$
F_3=\frac{(3m^4-k_0^2m^2r^2-2k_0^4r^4)(\Omega_0'\Omega_1-\Omega_0\Omega_1')}{m^2r(m^2+k_0^2r^2)\Omega_0^2}-\frac{\Omega_1''}{\Omega_0}+\frac{\Omega_1\Omega_0''}{\Omega_0^2}.\eqno{(\theequation e)}
$$

The standard adjoint theory for second order ordinary differential equations can be used to find the adjoint problem (\ref{wideadj}) with the operator
\begin{equation}
\begin{array}{c}\displaystyle
\mathscr{L}^\dag=\partial_{r}^2-\Big(\frac{3m^2+k_0^2r^2}{r(m^2+k_0^2r^2)}\Big){\partial_{r}}+\Big(-\frac{m^2}{r^2}-k_0^2+\frac{1}{r^2}+
\frac{6k_0^2m^2+4k_0^4r^2}{m^2(m^2+k_0^2r^2)} \vspace{.2cm}\\ \displaystyle
+
\frac{(-3m^4+k_0^2m^2r^2+2k_0^4r^4)\Omega'}{m^2r(m^2+k_0^2r^2)\Omega}
-\frac{\Omega''}{\Omega}\Big)-\Big(\frac{3m^2+k_0^2r^2}{r(m^2+k_0^2r^2)}\Big)'.
\end{array}
\end{equation}









\bibliographystyle{jfm}  
\bibliography{Reference}  

\end{document}